\documentclass[11pt]{article}
\usepackage{amsmath,amssymb}
\usepackage{graphics,epsfig}
\usepackage{graphicx}

\def\bea{\begin{eqnarray}}
\def\eea{\end{eqnarray}}
\def\be{\begin{equation}}
\def\ee{\end{equation}}

\newcommand{\ptl}{\partial}
\newcommand{\lal}{\langle}
\newcommand{\ral}{\rangle}
\newcommand{\lla}{\left\langle}
\newcommand{\rra}{\right\rangle}

\newcommand{\ovl}{\overline}

\newcommand{\vct}[1]{{\mbox {\boldmath $#1$}}}

\begin{document}

\begin{center}
{\bf \large A public turbulence database cluster and applications to study Lagrangian evolution
of velocity increments in turbulence}

\vskip 0.2cm
Yi Li$^{1,2}$, Eric Perlman$^3$, Minping Wan$^1$, Yunke Yang$^1$, Charles Meneveau$^1$, 
Randal Burns$^3,$ Shiyi Chen$^1$, Alexander Szalay$^4$, Gregory Eyink$^5$
\vskip 0.2cm
\noindent $^1$Department of Mechanical Engineering, The Johns Hopkins University 
(current address: $^2$Department of Applied Mathematics, University of Sheffield, UK), 
$^3$Department of Computer Science, $^4$Department of Physics and Astronomy, 
$^5$Department of Applied Mathematics and Statistics, The Johns Hopkins University, 
3400 North Charles Street, Baltimore, MD, 21218

\end{center}

\begin{abstract}
A public database system archiving a direct numerical simulation (DNS) data set of isotropic, 
forced turbulence is described in this paper. The data set consists of the DNS output on $1024^3$
spatial points and 1024 time-samples spanning about one large-scale turn-over timescale.  
This complete $1024^4$ space-time history of turbulence is accessible to users remotely through an 
interface that is based on the Web-services model.  Users may write and execute 
analysis programs on their host computers, while the programs make subroutine-like 
calls that request desired parts of the data over the network.  The users are thus 
able to perform numerical  experiments by accessing the 27 Terabytes of  DNS data 
using regular platforms such as laptops. The architecture of the database is explained, 
as are some of the locally defined functions, such as differentiation and interpolation. 
Test calculations are performed to illustrate the usage of the system and 
to verify the accuracy of the methods. The database is then used to analyse a 
dynamical model for small-scale intermittency in turbulence. Specifically, 
the dynamical effects of pressure and viscous terms on the Lagrangian evolution 
of velocity increments are evaluated using conditional averages calculated from the DNS data in the database. 
It is shown that these effects differ considerably among themselves and thus 
require different modeling strategies in Lagrangian models of velocity increments and intermittency.
\end{abstract}

\section{Introduction\label{sect:intro}}

With the advent of petascale computing, it will become possible to simulate turbulent flows that cover over 
4 orders of magnitude of spatial scales in each direction. 
For example, a simulation of turbulent flow that outputs 4 field variables on $O[(10^4)^3]$ spatial points, when stored on $10^4$ time-frames leads to datasets of 
about 160 petabytes. For more complex problems, such as turbulent advection of 
passive scalars \cite{SchumacherSreenivasan05,Yeung2002}  and 
magnetohydrodynamic turbulence \cite{Mininni2005}, the amount of data will increase even further. 
Such large datasets, however, create serious new challenges when attempting to translate 
the massive amounts of data into meaningful knowledge and discovery about turbulence. 
The natural answer to this challenge is to seek the application of ``database technology'' in 
Computational Fluid Dynamics (CFD) and turbulence research. Database technology more broadly 
has focused on seeking automated ways to identify patterns and reduced-order descriptions, 
develop machine learning, perform data mining, and so on, in order to reduce the amount of data 
to be processed and transmitted. One example of such projects is the Sloan Digital Sky 
Survey (SDSS) project (for a list of similar projects see the Appendix E of \cite{CI07}). 
The project aims at producing a comprehensive astronomical data archive, 
and providing astronomers the ability to explore the multi-terabyte  data interactively 
and remotely through the Internet \cite{Szalayetal00}. 
The data archive has led to over one thousand research articles published by researchers 
from all over the world. While there are several fields in which this approach has been embraced, 
the database approach has, to date, not been widely embraced by the CFD research community. 
For instance, in the area of Direct Numerical Simulations (DNS) of turbulent flows, 
existing efforts to share large datasets \cite{CINECA2007,Yeung2008,Hoyasetal2006} either 
(1) assume that the entire or most of the data are downloaded and the user 
must provide the computational resources for analysis of the large data sets, 
or (2) the user must establish accounts, privileges, and be granted access to large fractions of memory 
at a high-perforance computing (HPC) facility, typically where the data sets were generated and stored, 
in order to run analysis programs there. While valuable contributions, such approaches 
do not fully leverage existing database technologies, which should support scientists 
with flexible and user-friendly tools without requiring extensive 
initial efforts to set up specialized access to HPC facilities. Also, in order to 
accommodate the majority of researchers with more limited computational infrastructure, 
database approaches should allow scientists access to relevant parts of the data 
without having to download significant fractions of the data to their computers. 
The prevailing environment without such effective database support has hampered 
wide accessibility of large turbulence and CFD datasets. This problem is not expected to improve 
substantially even with faster network technologies. While data transfer is getting faster, 
the size of the ``top-ranked simulations'' is growing faster still. Thus, without developing new approaches, the top-ranked turbulence simulations may remain accessible only to a small subset of the research community.

As a step in the direction of applying database techniques to turbulence research, 
in this paper we describe the construction and application of a 27 Terabyte 
database cluster containing the $1024^4$ 
space-time history of forced isotropic turbulence obtained using DNS.  
The database archives a time series of velocity and pressure fields of isotropic stationary
turbulence. The data are obtained by DNS with $1024^3$ grid points in a periodic box.  
1024 time-samples spanning about one large-scale turn-over time scale are stored.
A user interface is layered on top of the database, so that the users can access the data via
the Internet. A function set implemented in the database allows flexible {\sl in situ}
data processing for basic operations such as differentiation and interpolation. 
While some of the technical details about the design of the database have been described in 
\cite{Perlmanetal07}, in this paper we give a further description, 
with emphasis on how it may be used for turbulence research. 
We also provide a description of the functions available to the user.  
Then several test cases are presented to illustrate the usage of the database. 
The tool is then applied to the analysis of a Lagrangian model of turbulence 
small-scale intermittency proposed recently \cite{LiMeneveau05,LiMeneveau06}.

Turbulence intermittency refers to the infrequent but strong fluctuations in the
parameters characterizing small-scale turbulent motions, such as velocity
increments, velocity gradients, and dissipation rates. Due to small-scale
intermittency, it is observed in experiments and simulations that
the probability density functions (PDF) of the small scale quantities
develop exponential or stretched exponential tails. Also, the scaling
exponents of the moments of the velocity increments deviate
significantly from the prediction of the classical Kolmogorov theory
\cite{Kolmogorov41a, Frisch95, Pope00}. Thus the quantitative prediction of small-scale intermittency
behavior has been one of the major fundamental challenges in turbulence research.
Substantial progresses have been made in the phenomenological
modeling and related `toy models' of small scale intermittency
\cite{Benzietal84,MeneveauSreenivasan87, Kraichnan90, SheLeveque94,
Falkovichetal01, Biferale03}. However, these models generally
made little direct connection with the Navier-Stokes equations.
A recent development has been to couple the dynamics of velocity
gradients with the Lagrangian evolution of the small scale geometrical features
of turbulence. Geometrical information has been invoked in the  tetrad model \cite{Chertkovetal99},
in the material deformation tensor evolution\cite{JeongGirimaji03,
ChevillardMeneveau06}, and material line elements \cite{LiMeneveau05,
LiMeneveau06}. In Refs. \cite{LiMeneveau05, LiMeneveau06}, the
velocity increments defined over a line segment advected by a
turbulent velocity field have been studied. A simple dynamical system
describing the Lagrangian evolution of the velocity increments was
derived. In this so-called ``advected delta-vee" system, the
nonlinear self-interaction terms in the Navier-Stokes equations are
reduced to some simple closed terms. Neglecting the unclosed terms,
the truncated system was shown to reproduce several important
trends concerning intermittency, such as the skewness and elongated tails in the PDFs of
the velocity increments.

However, since several important pressure and viscous terms were neglected in the ``advected delta-vee" system, the system does not approach a stationary state and instead diverges towards unphysical
behavior at large times. As concluded in \cite{LiMeneveau06},
quantitative prediction of intermittency requires the modeling of the
unclosed terms. In this paper,
the terms are analyzed using the DNS data in the turbulence database.
Following the statistical approach of \cite{VanderBosetal02},
we consider the effects of the pressure and viscous terms
based on the probability fluxes they produce in the equation for the joint probability
distribution function (PDF) of the velocity increments.

The paper is organized as follows: In \S \ref{sect:arch}, some design principles of
the database are described, and the dataset is also documented.
Several standard calculations are conducted in \S
\ref{sect:dbtests}, which include the one-dimensional energy spectra of the
velocity fields \cite{Pope00}, the PDFs
of elements of the velocity gradients $A_{ij}=\ptl u_i/\ptl x_j$, and the joint
PDF of the two independent invariants of the velocity gradient
$Q=-A_{ij} A_{ji}/2$ and $R=-A_{ij}A_{jk}A_{ki}/3$ \cite{Cantwell92}. The statistical descriptions
are obtained by running programs on local desktop computers, while obtaining the data from the database remotely. The analysis of the advected delta-vee system is presented in \S \ref{sect:deltavee}.
The conclusions are given in \S \ref{sect:conclusions}.

\section{Architecture of the database cluster \label{sect:arch}}
\begin{figure} [ht]
\centering
\includegraphics[width=0.6\linewidth]{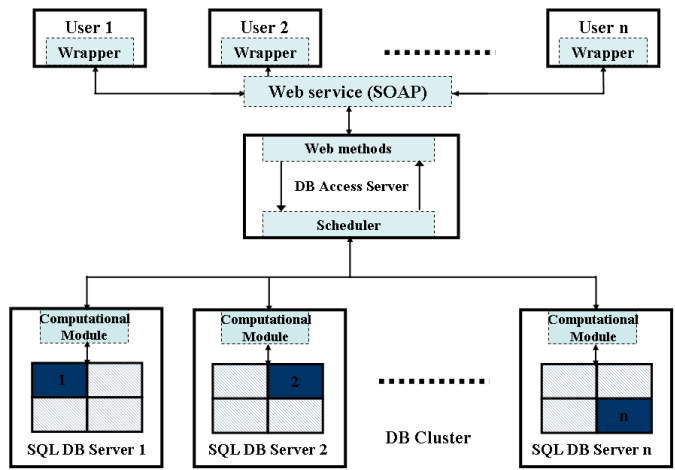}
\caption{\label{fig:DatabaseArch} The architecture of the turbulence database
cluster.}
\end{figure}
From the point of view of a user, it is
desirable that a database cluster should, among others: (1) be available to the public by
allowing remote access through the network; (2) provide flexible and
comprehensive data processing functionalities inside the database so
as to reduce data download; (3) be compatible with different
platforms; (4) be compatible with the programs and tools that have
been developed and used by the researchers in the past; (5) be efficient.
With these requirements in mind, a cluster of database servers
has been designed according to the diagram shown in Fig.
\ref{fig:DatabaseArch}. A time series of $1024^3$ DNS data of
isotropic, steady state turbulence is stored in the database
cluster. A total of 1024 time steps, or $27$ terabytes of data, have been
loaded. The data are partitioned spatially and placed on different database
nodes in the cluster.  Most of the nodes have two dual core AMD Opteron
processors with 4GB memory, while some have two Intel Xeon quad-core processors. 
The operating system is 64-bit Microsoft
Windows 2003 Server. The dataset is managed by Microsoft SQL Sever 2005 (64 bit). For more details
see \cite{Perlmanetal07}.

The users connect to the database via the Internet (at http://turbulence.pha.jhu.edu)
and access the data through a data access server. The communication between the
users and the data access server adopts the Web service model. The data access server 
is also the head-node of the database cluster. It communicates with 
other database servers (nodes) in the cluster over high-speed local area network.
The data access server is a mediator: it receives requests from the users,
breaks the request into parts and sends the parts to the individual
database nodes. The majority of the calculations are done in the
Computational Module in each node, embodying the ``move the
program to the data" principle in scientific database design
\cite{Szalayetal00}. High efficiency is achieved through a high degree of
parallelism. The details of the turbulence dataset and details of several components of the
database cluster are described in the following several sub-sections.

\subsection{Turbulence data set}
The data is from a DNS of forced isotropic turbulence on a $1024^3$ 
periodic grid in a $[0,2\pi]^3$-domain, using a pseudo-spectral parallel code.
Time integration of the viscous term is done analytically using the exact integrating factor.
The other terms are integrated using a second-order Adams-Bashforth scheme and the nonlinear 
term is written in vorticity form \cite{CaoChen99}. The simulation is 
de-aliased using phase-shift and $2\sqrt{2}/3$ spherical truncation \cite{PattersonOrszag71}, 
so that the effective maximum wavenumber is about $k_{\rm max}=1024\sqrt{2}/3\approx 482$. 
Energy is injected by keeping constant the total energy in modes such that their 
wave-number magnitude is less or equal to 2. The simulation time-step is $\Delta t = 0.0002$.
After the simulation has reached a statistical stationary state, $1024$ frames of data, 
which includes the 3 components of the velocity vector and the pressure, 
are generated in physical space and ingested into the database. The data are stored at every 10 DNS time-setps, 
i.e. the samples are stored at time-step $\delta t = 0.002$.  The duration 
of the stored data is $1024 \times 0.002 = 2.048$, i.e. about one large-eddy turnover time since 
$T=L/u' \approx 2.02$ ($L$ is the integral scale).  The parameters of the 
simulation are given in Table \ref{tab:dnsparam}. The total kinetic energy, 
dissipation rate, and energy spectra are computed by averaging in time between $t=0$ and $t=2.048$.
\begin{table}[ht]
\renewcommand{\arraystretch}{1.2}
\center
\begin{tabular}{l cl}
\hline\hline
Resolution, $N$ & \ & 1024 \\
Viscosity, $\nu$ & \ & 0.000185\\
Time step of DNS, $\Delta t$ &\ & 0.0002\\
Time interval between stored data sets, $\delta t$ &\ & 0.002 \\
Total kinetic energy, 
$E_{\rm tot} = \langle \sum \limits_{\bf k} \frac{1}{2} \hat{\bf u}\cdot \hat{\bf u}^*\rangle_{\rm t}$ & \ & 0.695\\
Mean dissipation rate, $  \epsilon  =\langle \sum \limits_{\bf k} \nu k^2 \hat{\bf u}\cdot \hat{\bf u}^*\rangle_{\rm t}$ &\ & 0.0928\\
r.m.s. velocity fluctuation, $u'=\sqrt{\frac{2}{3}E_{\rm tot}}$ & \ & 0.681\\
Taylor micro length scale, $\lambda=\sqrt{ 15 \nu u'^2/\epsilon}$ &\  & 0.118\\
Taylor micro-scale Reynolds number, $Re_{\lambda} =  u'\lambda/\nu$ &\ & 433\\
Kolmogorov length scale, $\eta_K=(\nu3/\epsilon)^{1/4}$ & \ & 0.00287\\
Kolmogorov time scale, $\tau_K=(\nu/\epsilon)^{1/2}$ & \ & 0.0446\\
Integral length scale, $L=\frac{\pi}{2u'^2} \int \frac{E(k)}{k} dk$ & \ & 1.376\\
Integral time scale, $T=\frac{L}{u'}$ & \ & 2.02\\

\hline \hline\\
\end{tabular}
\label{tab:dnsparam}
\caption{The parameters of the DNS data in the turbulence database.}
\end{table}

\begin{figure} [h]
\centering
\includegraphics[width=0.6\linewidth]{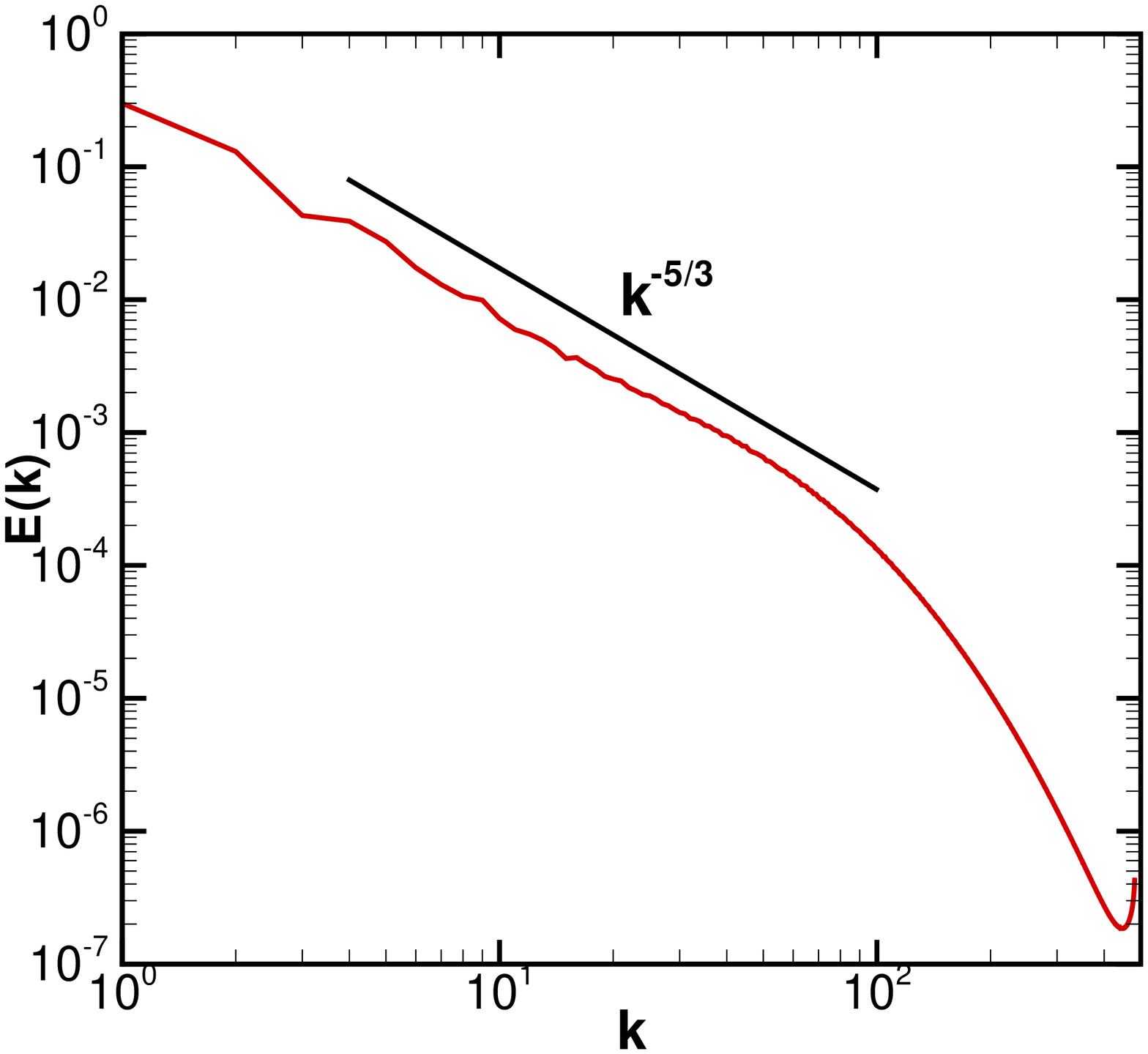}
\caption{Radial kinetic energy spectrum, averaged in time between $t=0$ and $2.048$.}
\label{fig:radialspectrum}
\end{figure}

\begin{figure} [h]
\centering
\includegraphics[width=0.6\linewidth]{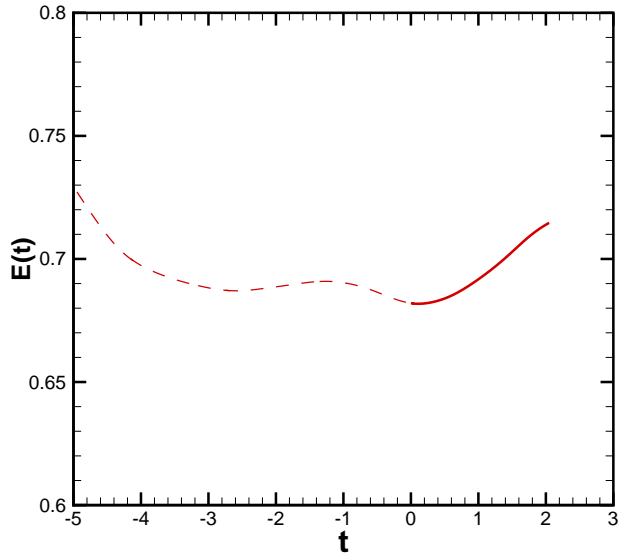}
\caption{Total kinetic energy as function of time. Dashed line is times before ingestion 
into database. Data corresponding to the database is show using solid line between $t=0$ and $2.048$.}
\label{fig:historyE}
\end{figure}

\begin{figure} [h]
\centering
\includegraphics[width=0.6\linewidth]{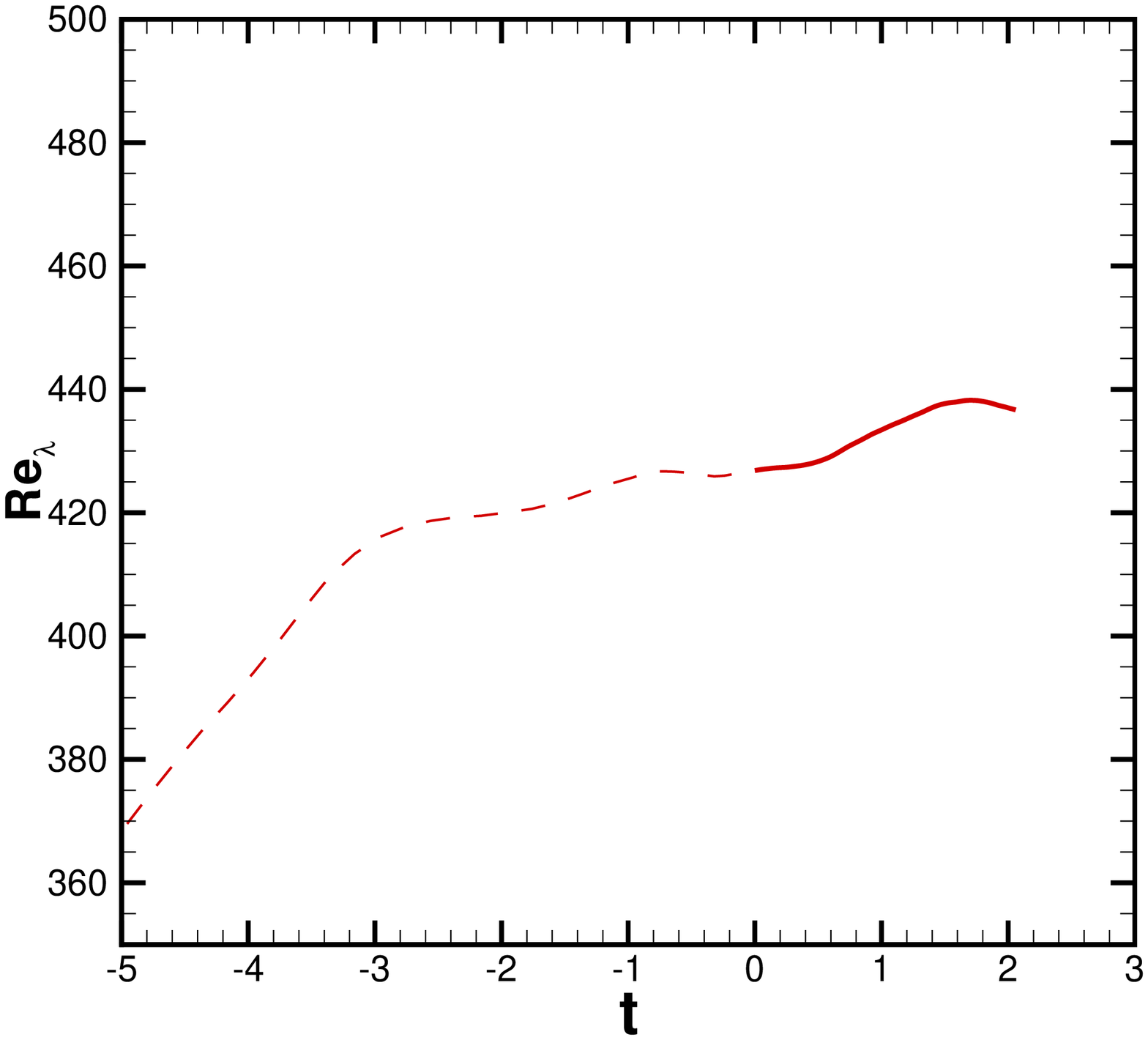}
\caption{Taylor micro-scale Reynolds number as function of time. 
Dashed line is times before ingestion into database. 
Data corresponding to the database is show using solid line between $t=0$ and $2.048$.}
\label{fig:historyRe}
\end{figure}

Figure \ref{fig:radialspectrum} shows the radial energy spectrum computed from the simulation and averaged between   
$t=0$ and $t=2.048$. Figures \ref{fig:historyE} and \ref{fig:historyRe} show time-series of 
the total turbulent kinetic energy and Taylor-scale based Reynolds numbers 
starting near the simulation initial condition. The values corresponding to the data 
in the database are for $t>0$ and are shown in solid line portions.

\subsection{User interface}
The Web service model is adopted to implement the communication between
the users and the database. The Web service technique makes possible the
automatic interaction between the database and a computational program running on a
user's desktop or laptop computer. The program can fetch data
from the database whenever  data are needed, while performing
further data analysis on the local machine between data requests.
To realize this functionality, the user interface includes two parts: the
Web-service methods implemented on the database access server and the
client-side wrapper interface. Web-service methods are based on the
standard protocol SOAP (Simple Object Access Protocol) \cite{SOAP}.
They can be easily called from some modern programming languages such as 
Java or C\#.  Unfortunately, other languages need additional libraries 
to use Web-services.  For example while recent versions of MATLAB offer 
integrated Web-service support, it is still necessary to provide a 
wrapper function to perform complex data type conversions.  
For FORTRAN and C codes running in UNIX, Linux or MacOS, we have developed
and provided FORTRAN and C wrapper interfaces to the
gSOAP library \cite{gSOAP}, which establishes the communication
between the client code and the Web-service methods.

A list of Web-service methods are implemented on the data access server
\cite{turbweb}. These methods provide the entry points to the
processing functions implemented on the database servers (the
``Computational module" in Fig. \ref{fig:DatabaseArch}, will be
explained below). The data access server has the knowledge of the data
partition among the database servers. When a request is received
from the user, the data access server find out on which database servers the
requested data is stored, and then break the request into parts and
send each part to the corresponding server. Upon
receiving the requests, the Computational Modules on the database
servers are invoked to perform necessary computations and return the
requested data to the Web-service methods on the data access server (see
Fig. \ref{fig:DatabaseArch}). The Web-service methods then communicate the
data back to user programs.

\subsection{Processing function set}
A set of processing functions are implemented as stored
procedures on the database servers, referred to as Computational Module 
in Fig. \ref{fig:DatabaseArch}. Microsoft SQL Server's Common Language Runtime
(CLR) \cite{LamThai03} integration allows these stored procedures
programmed in C\#. They are intended to provide for the users a
comprehensive and yet compact set of tools to process the data
in the database.  In turbulence research,  the questions being asked and
techniques used in data analysis are highly specific to the client
and often vary from one client to another. Therefore, the design of
these functions is not straightforward. One has to consider the
trade-off between generality and efficiency. From our experience, we
consider the most basic common tasks in data analysis are to
calculate some basic flow parameters on a set of spatial locations.
The basic functions supported thus include evaluating the velocity $u_i$ and pressure $p$ 
on arbitrary spatial locations, and also their spatial differentiation for first and
second derivatives, as well as spatial and temporal interpolation. Specifically,
the functions allow evaluating the full velocity gradient tensor $A_{ij} = \partial u_i/\partial x_j$
as well as the pressure gradient $\partial p/ \partial x_i$.
For second derivatives, the pressure Hessian $\partial^2 p/\partial x_i \partial x_j$,
Laplacian of velocity $\nabla^2 u_i$,
and full velocity Hessian, $\partial^2 u_i/\partial x_p \partial x_q$, are supported.
Due to the need for spatially localized differentiation stencils, derivatives
are evaluated using centered finite differencing of 4th, 6th and 8th orders.
Spatial interpolation is performed using Lagrange polynomial interpolation of
various orders (also 4th, 6th and 8th order).  Temporal interpolation is
done using Cubic Hermite Interpolation. Details of these functions are provided
in Appendix B, C, and D.

\subsection{Data indexing, partition, and workload schedule}

\begin{figure} [ht]
\center \includegraphics[width=0.5\linewidth]{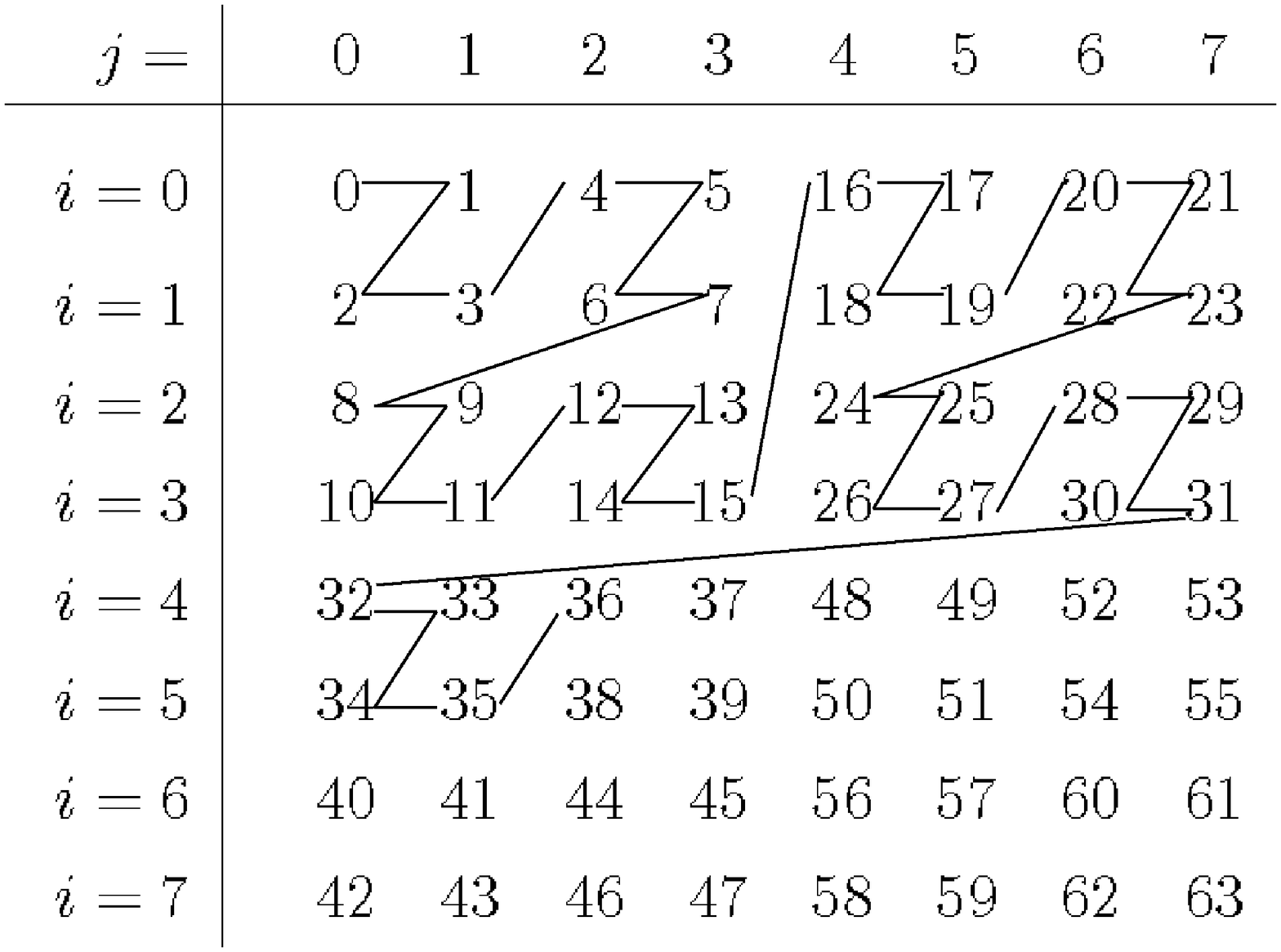}
\caption{\label{fig:z-curve} Table of Z-indices of a
two-dimensional $8\times 8$ grid. $i$ and $j$ are the indices of the
grid points along $x$ and $y$ directions, respectively. The values in
the table are the Z-indices of the grid points calculated using
bit-interleaving. The Z-order curve connects the grid points
according to Z-order, i.e., in the order of increasing Z-indices. The
same order is used to index the data in the database.}
\end{figure}

Indexing the spatial data consists of mapping the three dimensional
spatial data onto a one dimensional array. Maintaining good locality
of the spatial data is one of the favorable attributes of the
mapping. That is, if two points are close to each other in the
3D space, their positions in the one dimensional array should also be
close. This is clearly an advantage for applications accessing
coherent regions in the physical space, such as structure
identification and filtering. In our system, the data is indexed
according to the Z-order (also known as Morton-order)\cite{Samet06}, for it maintains fairly
good locality and is simple to implement and
calculate. The Z-index of each grid point in the computational mesh
is calculated from the three-dimensional index of the point using
bit-interleaving \cite{Perlmanetal07}. The concepts of Z-order and
Z-index are illustrated in Fig. \ref{fig:z-curve} with a
two-dimensional grid. For our $1024^3$ DNS data,
we need $30$ bits to index all the points. The Z-index of each point
has $30$ bits in its binary representation.

The spatial partition and organization of the data is also based on
the Z-order of the data. To optimize I/O operations, we divide the
whole data set into ``atoms" of a fixed size (see below). 
The atom is
the fundamental unit of I/O operations, i.e., whenever some data is
needed, a whole atom is read into the memory. 
The atoms are stored on disk and indexed using a standard database B+-tree index.
The B+-tree index of an atom is given by the six
common middle bits of the binary representations of the Z-indices of
the data points within the atom (see \cite{Perlmanetal07} for
details). Using these bits as the search key in the B+-tree ensures
that the atoms that are contiguous in the Z-order are also
contiguous in the index. Given the good
locality of the Z-order, this data structure thus supports fast
access to contiguous region in the physical space. Similarly, the
allocation of the atoms on different data servers is also determined
by the Z-indices of the data points.

\begin{figure} [ht]
\centering
\includegraphics[width=0.7\linewidth]{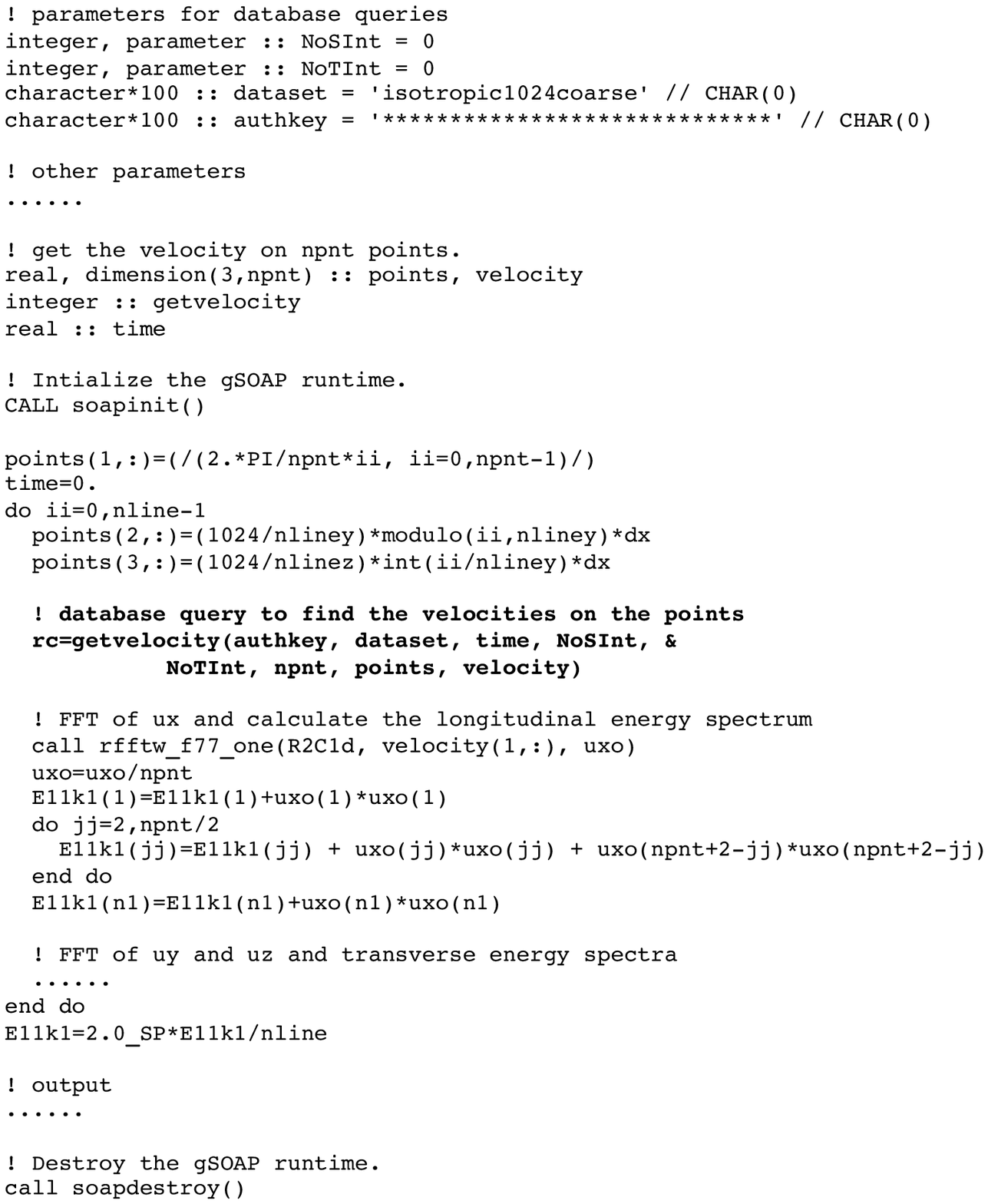}
\caption{Snippet of the FORTRAN code running on local user machine.
Bold font highlights the lines invoking the Web-service
method. The authkey has been intentionally marked out.
\label{fig:FORcode}}
\end{figure}
\begin{figure} [ht]
        \centering
		\includegraphics[width=0.7\linewidth]{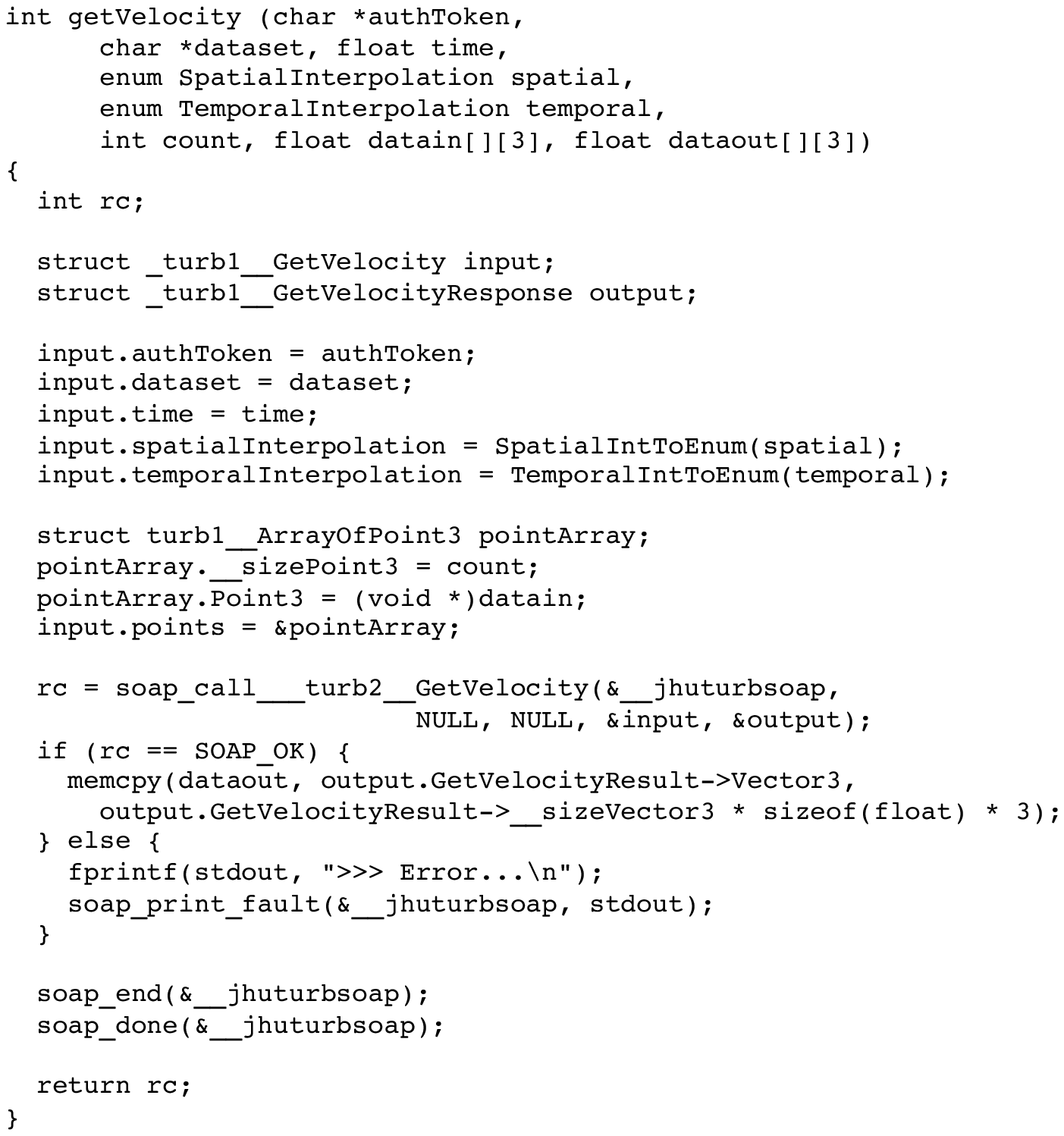}
        \caption{The C wrapper interface for the FORTRAN code. It invokes
        the gSOAP libraries to invoke the Web-service method. \label{fig:turblib}}
\end{figure}
       
The size of the atoms is chosen based on two considerations. We have found that
using an atom of size $64^3$ data points as the fundamental I/O unit 
is the most efficient because it balances the memory
and disk performance \cite{Perlmanetal07} in our application. However, as 
we mentioned before, a frequent operation required by the
processing functions is interpolation. For 8th order
interpolation, a blob of $8^3$ data points need to be accessed to
obtain the interpolated value at one point. Given the high frequency
of this operation, it is highly desirable to avoid data transfer
between different data servers during this operation, which would
incur significant overhead. This is done by ``edge replication" on the atoms. Specically, we divide 
the data space into $(1024/64)^3 = 16^3$ cubes of size $64^3$. An atom 
contains a cube in its center, and an edge of length 4 on each side of the
cube. Each atom thus has $72^3$ data points, with edges replicated
from the neighoring atoms.   This way, it is
guaranteed that each interpolation operation can be finished with no
more than one atom being read \cite{Perlmanetal07}.

The Z-order indexing scheme is also used to schedule workload. As
introduced above, the generic job request would be finding the
values of some parameter on an array of spatial locations. To
calculate the parameters at a point, the data in a small region
around the point are accessed and need to be read from the hard
drive. To minimize the amount of data reading, the spatial
locations, and thus the data queries, are grouped according to the
Z-order. Combined with data caching, where an atom of data is kept
in the memory until another atom has to be loaded, this scheduling
guarantees the same data atom is loaded into memory only once per job request.

\section{Preliminary numerical tests\label{sect:dbtests}}
While loading the data onto the database cluster a number 
of point-by-point checks are made.
In order to illustrate the usage of the database system, 
and also as a further check against, for example, the potential errors 
occurred during loading the data into the database, three basic groups of quantities are calculated using
the database. The first group is the longitudinal and transverse one-dimensional (1D) energy spectra.
The longitudinal spectrum is defined as
$E_{11}(k_1)=\langle \hat{u}_1(k_1^*)\hat{u}_1(k_1)\rangle$, where $\hat{u}_1(k_1)$ is the
one-dimensional Fourier transformation of the $x$ component of the velocity along the $x$ direction.
The average is taken over the $(y,z)$ plane.
Similarly, the two transverse spectra are defined as $E_{22}(k_1)=\langle \hat{u}_2(k_1^*)\hat{u}_2(k_1)\rangle$ and $E_{33}(k_1)=\langle \hat{u}_3(k_1^*)\hat{u}_3(k_1)\rangle$, where $\hat{u}_2$ and $\hat{u}_3$ are
one-dimensional Fourier transformations of the $y$ and $z$ components of the velocity, respectively.
The second group of tests is the PDFs of the
longitudinal and transverse velocity gradients.
The longitudinal velocity gradients are the components of $A_{ij}$ when $i=j$,
while the others are considered transverse ones.
The third one is the joint PDF of the two tensor invariants of the velocity gradient
$R\equiv -A_{ij}A_{jk}A_{ki}/3$ and $Q=-A_{ij}A_{ji}/2$\cite{Cantwell92}.
The tear-drop shape of the joint PDF of these quantities 
is a well-known feature of turbulence \cite{Cantwell92,Chertkovetal99,VanderBosetal02,ChevillardMeneveau06}.
These three basic groups are evaluated in two separate ways and compared:
The first is  in the traditional way, which we will call ``in-core" analysis.  
The in-core analysis is done on the cluster where the data are generated, 
and the raw data are the Fourier components of the velocity fields  
before they are transformed and loaded into the database.  
The velocity gradients are calculated in the Fourier space using spectral derivatives.
The second approach uses the data stored in the database and uses database queries. 
In particular, the velocity gradients are obtained using the $6$th 
order centered finite differencing option (see Appendix B). 
Interpolation of the velocity is done using 6th-order Lagrange polynomial option when necessary 
(see Appendix C). In order to remove possible differences resulting from statistical sampling, data on 
exactly the same spatial locations are used in both methods.

The 1D spectra are calculated based on the velocity along particular lines across the domain. These are 
obtained by invoking the Web-service method \verb=GetVelocity=. We
take $512$ grid lines along the $x$ direction at $32\times 16$ $(y,z)$ locations.
The $32\times 16$ locations are distributed uniformly on the grids along both $y$ and $z$ directions,
i.e., $32$ locations along $y$ direction and $16$ locations along $z$ direction.
In each database query, the velocity vectors on the $1024$ grid points on one of the lines are fetched from the
database. Therefore there are $512$ database queries in total, each fetching $1024$ points.
For the analysis in this paper, only data at a single time, at $t=0$, is used.
The one-dimensional FFT of the velocity data along each
line is performed in the program running on the local user machine.
The energy spectra are obtained by averaging over all the lines. The
FORTRAN code snippet in Fig. \ref{fig:FORcode}, taken from the
program running on the local user machine, shows how the database is
typically used in an application. As is highlighted in the figure
with bold font, the Web-service method is invoked from the computational
program through a wrapper function having the same name
\verb=GetVelocity=, as if it is another function implemented on the
same computer. The wrapper function takes the time, the number
of points, the order of Lagrange interpolation to be used, and the
$x$, $y$, and $z$ coordinates of the points as input parameters. The
three velocity components on the requested points are returned. The
C wrapper function, shown in Fig. \ref{fig:turblib}, invokes the
gSOAP library to invoke the Web-service method implemented on the data
access server.
\begin{figure} [ht]
        \centering
        \includegraphics[width=0.7\linewidth]{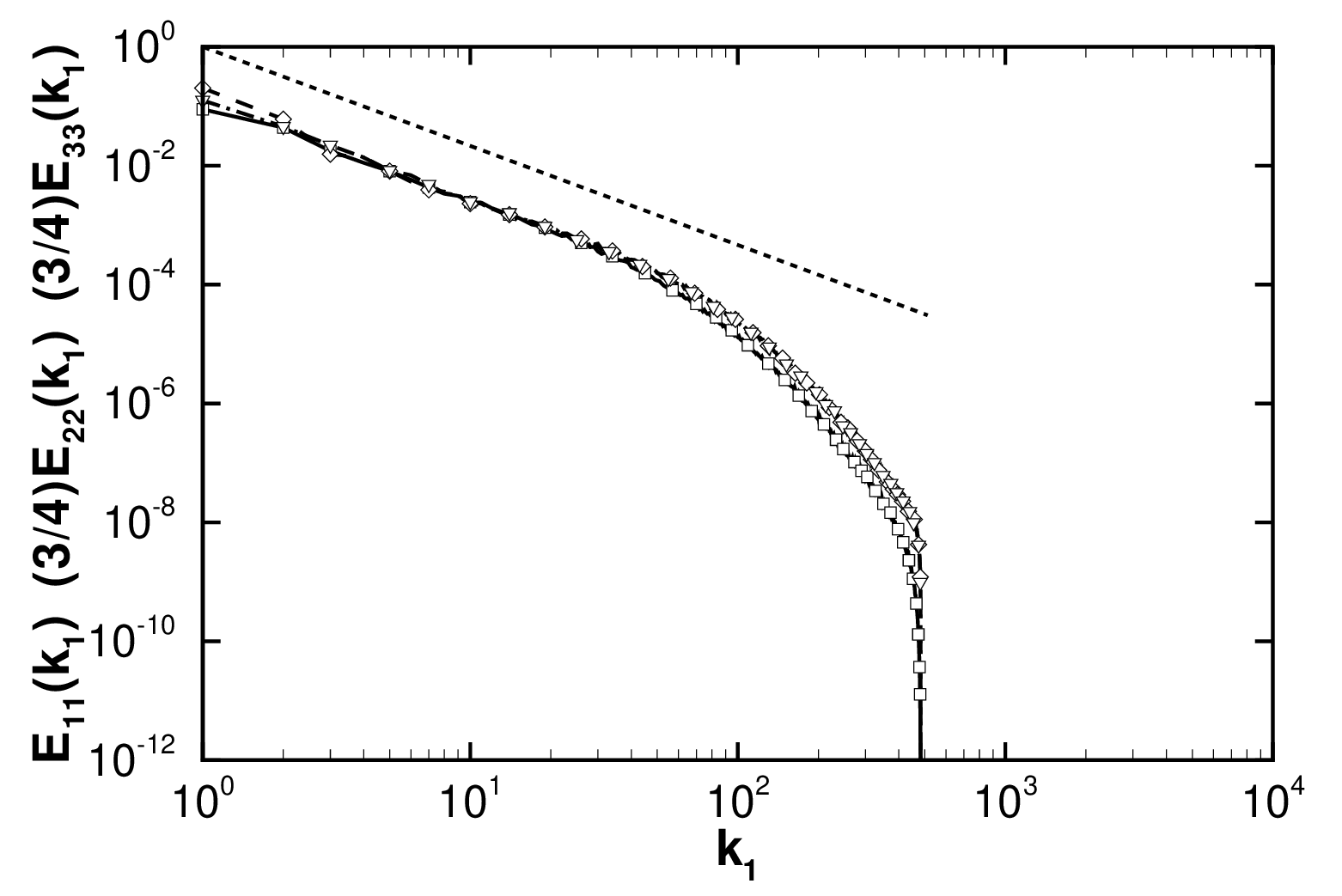}
        \caption{One-dimensional energy spectra calculated through database
        queries (lines) and calculated in-core (symbols). Solid line and
        squares: $E_{11}(k_1)$, dashed and diamonds: $(3/4)E_{22}(k_1)$,
        dash-dotted and gradients: $(3/4)E_{33}(k_1)$. Thin dashed line has
        slope $-5/3$.\label{fig:Ek1d_db}}
\end{figure}
Notice that the data points we are using are all on the grids.
        Because there is no interpolation error for the values of the
        velocity on these points, this calculation could be compared with
        the in-core calculation and serves as a
        low-level check for the correctness of the data loading process.
        The result is shown in Fig. \ref{fig:Ek1d_db}. The results calculated from the
        database are equal to those calculated in-core, suggesting that
        there was no error in loading the data. 
        
        The spectra show a narrow
        inertial range, in which the longitudinal and the rescaled
        transverse spectra overlap. The rescaled transverse spectra fall slightly
        above the longitudinal one towards the viscous range. The two
        transverse spectra are identical to each other except that slight
        difference can be seen at the lowest wavenumber. These features are
        all consistent with other DNS and experimental results (see, e.g.,
        \cite{Kangetal03}). As a result of dealiasing, the spectra extend
        only to a wavenumber around 482, as noted before.
        \begin{figure} [ht]
        \centering
        \includegraphics[width=0.7\linewidth]{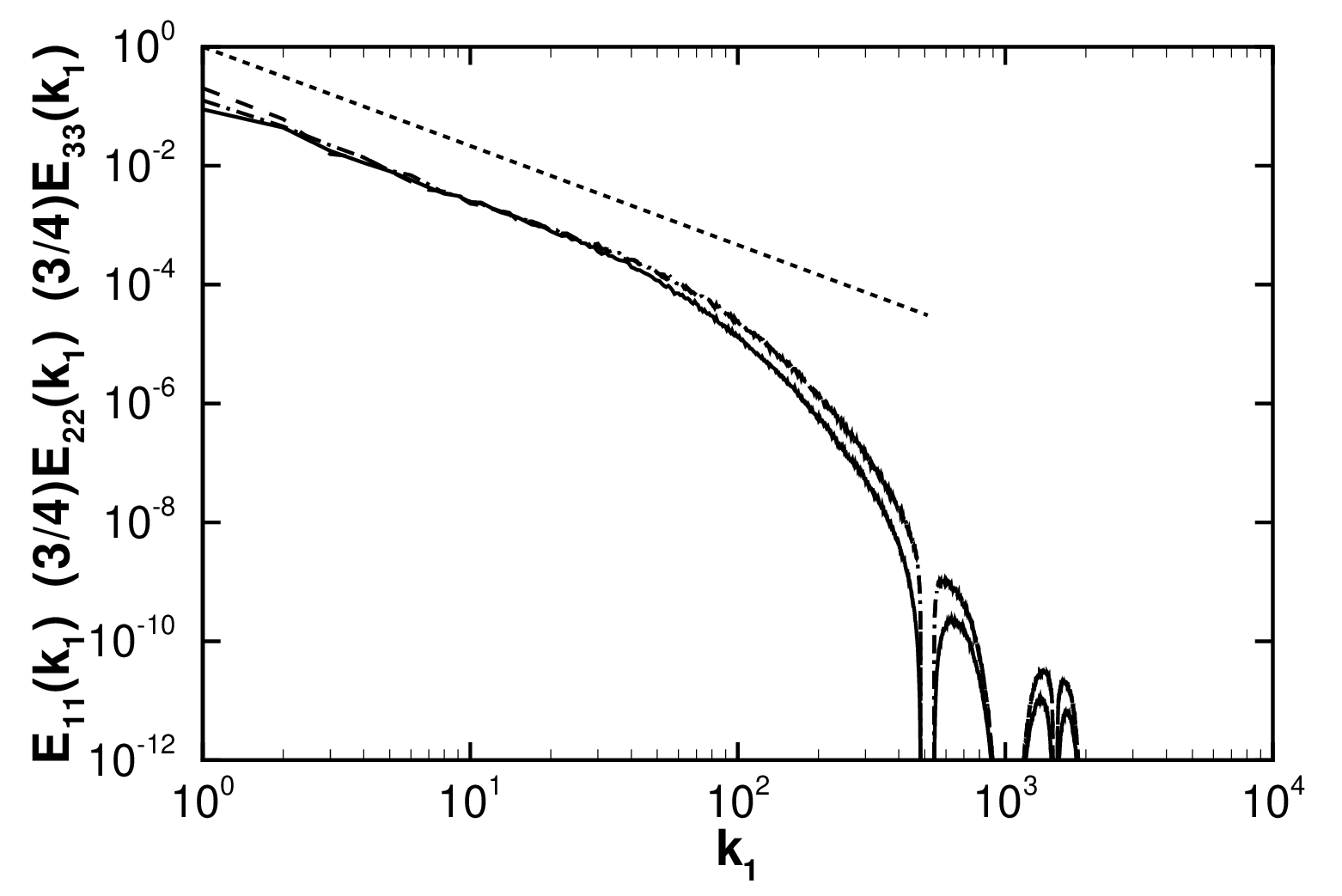}
        \caption{One-dimensional energy spectra calculated through database
        queries using $4096$ data points on each line with spatial interpolation.
           Line patterns same as in Fig. \ref{fig:Ek1d_db}. \label{fig:Ek1d_db_4096} }
        \end{figure}

        To illustrate the effects of Lagrange polynomial interpolation, another case
        with 512 lines and $4096$ points on each line is also calculated.
        The lines are chosen to be the same as in the previous case. The 4096
        points are uniformly distributed on each line, and one in every four
        points falls on the grids. The $6$th order Lagrange interpolation
        scheme is used. As is shown in Fig. \ref{fig:Ek1d_db_4096}, the
        range of the spectra extends beyond the maximal resolved wavenumber
        in the simulation (about 482) to a value of around 2000. The
        oscillating lobes observed between wavenumbers 482 and 2000 are 
	the spectral signature of the Lagrange interpolant polynomials.
        As a consequence of the smoothness of the interpolants, very little energy is
          contained in high wavenumber lobes,
           as one can see by comparing Fig.
        \ref{fig:Ek1d_db} and \ref{fig:Ek1d_db_4096}.

        The PDFs of the velocity gradients and the joint PDF of $R$ and $Q$
        are calculated in a similar way, except that the Web-service method invoked
        is \verb=GetVelocityGradient=, instead of \verb=GetVelocity=. The
        database queries return values for each element of the  
        velocity gradient tensor, at every point requested. 
        The velocity gradient tensor elements are calculated inside the database. In this case
        data on $256^3$, or about 17 million, spatial locations are requested. 
	 The points are again on the grids, and are equally spaced along each direction.

\begin{figure} [ht]
        \centering
        \includegraphics[width=0.7\linewidth]{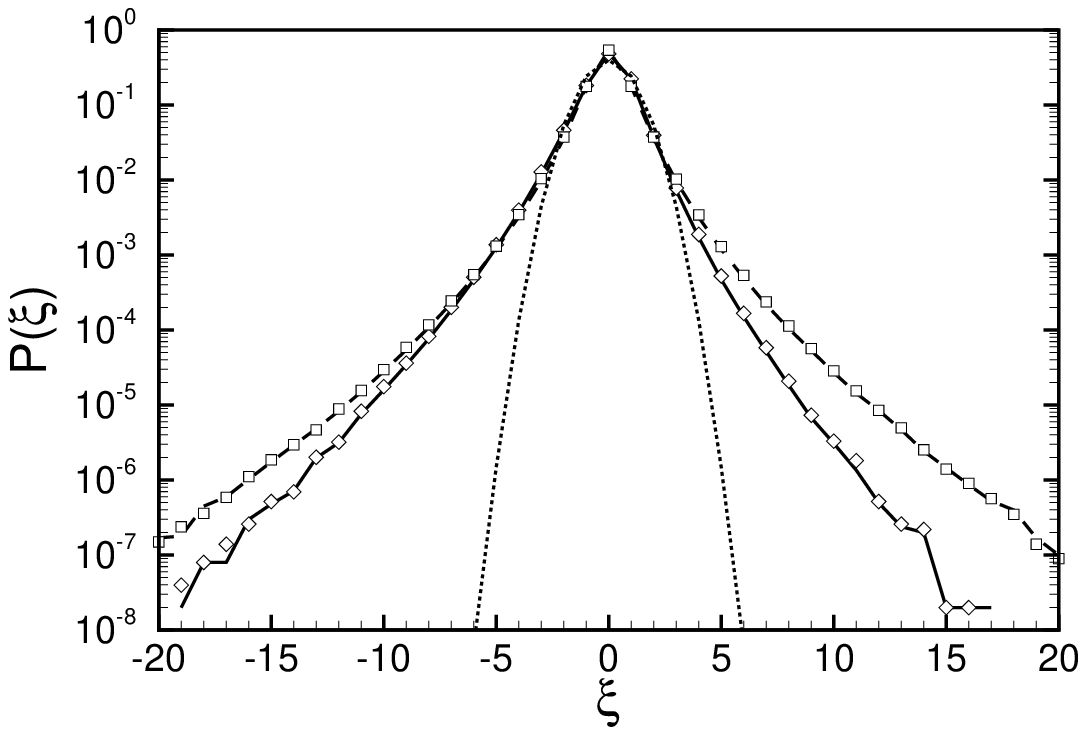}
        \caption{PDF of longitudinal (solid line and diamonds) and
        transverse (dashed line and squares) velocity gradients, normalized
        by rms values. Lines are calculated from database queries and
        symbols are from in-core calculations. The dotted line is Gaussian
        distribution\label{fig:pdudx_db}.}
        \end{figure}

\begin{figure} [ht]
        \centering
        \includegraphics[width=0.7\linewidth]{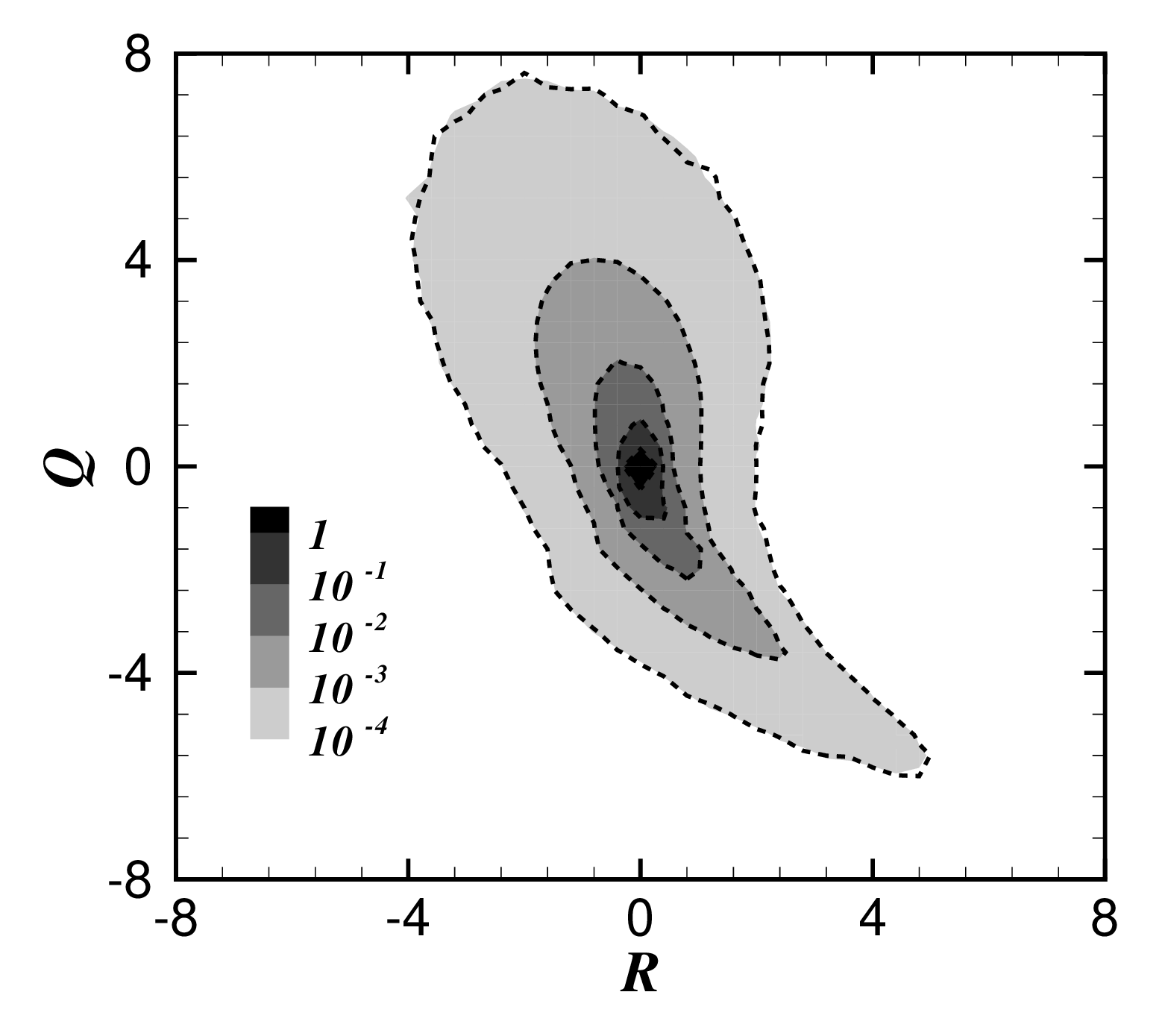}
        \caption{The contour plot of the joint PDF of $R$ and $Q$. Dashed
        lines: in-core calculations. Grey-scaled contours: database queries,
        in which $R$ and $Q$ are calculated on the local machine with the
        velocity gradient data fetched from the database.\label{fig:qr_db}
        }
        \end{figure}

        The results are plotted in Figs. \ref{fig:pdudx_db} and
        \ref{fig:qr_db}. Also plotted are the results calculated in-core, in
        which the velocity derivatives are calculated from the Fourier
        components of the velocity field, and then transformed into physical
        space. The figure shows that there is very good agreement between
        the results. The minor differences are due to slight differences in 
		the 6th-order finite difference and the spectral evaluations of the gradients. 
		The results are also in good agreement with the
        results reported in the literature (see, for example,
        \cite{Gotohetal02} and \cite{VanderBosetal02}).


        The time taken to calculate the energy spectra is about 16 minutes for
        $512\times 1024$ points, and 26 minutes for $512\times 4096$ points.
        These numbers are quite encouraging
        for an initial implementation that has much room for additional
        optimizations. 
        
\begin{figure} [ht]
        \centering
       \includegraphics[width=0.5\linewidth]{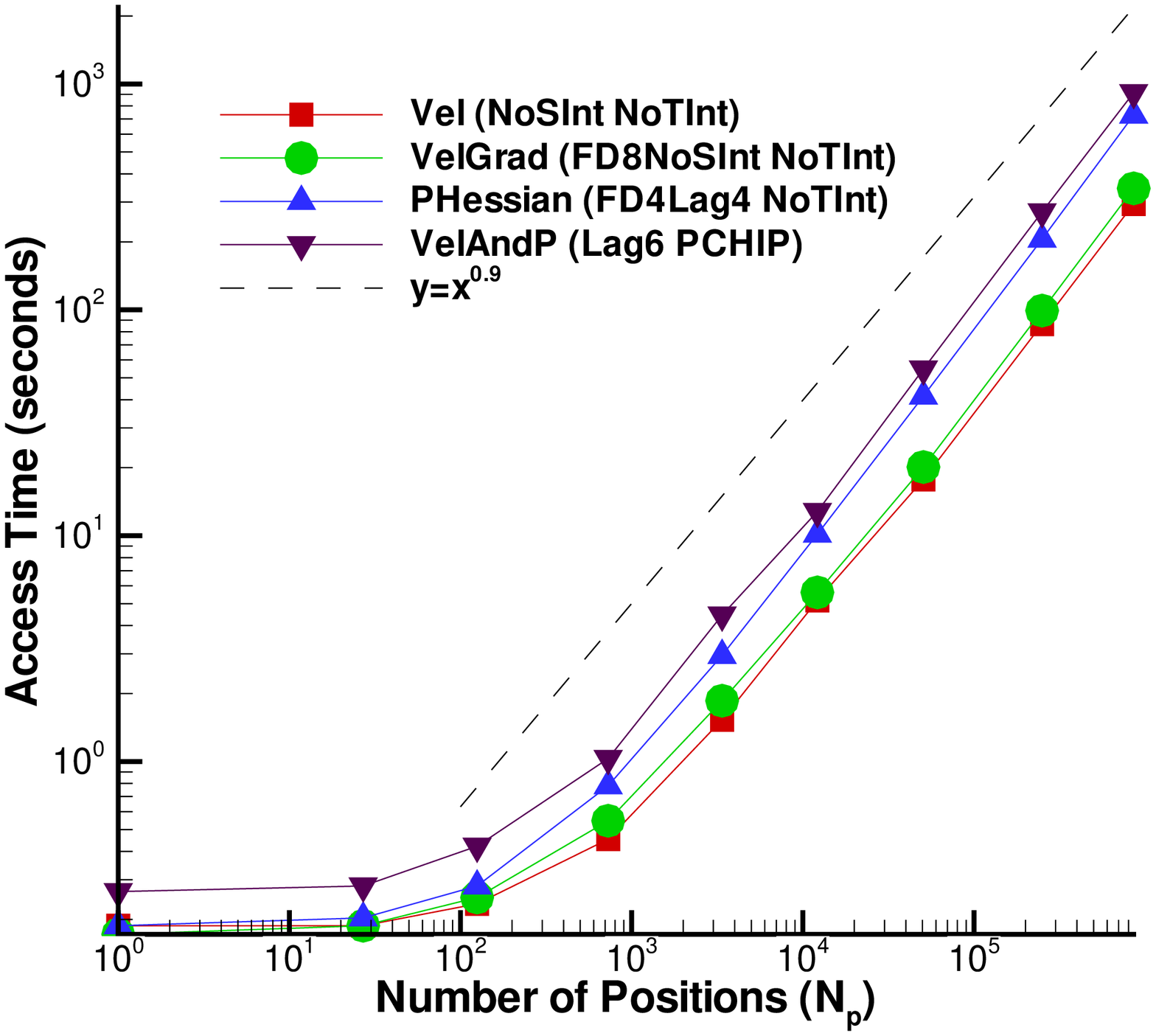}
        \caption{Access times (in seconds) measured as function of request size 
	(number of points $N_p$), when the $N_p$ points are arranged on a cubic lattice of size
	$\sim N_p^{1/3}$. Different symbols denote different request types, including GetVelocity, 
	GetVelocityGradient, GetPressureHessian, and  GetVelocityAndPressure. 
	Different combinations of order of interpolation and differentiation are also tested as indicated in the figure. 
	\label{fig-timesa}}
        \end{figure}

\begin{figure} [ht]
        \centering
        \includegraphics[width=0.5\linewidth]{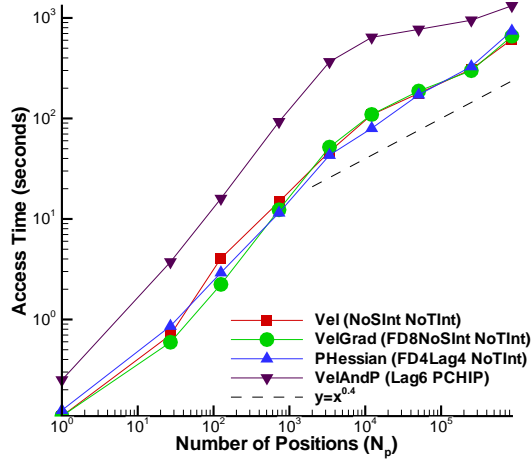}
        \caption{Access times (in seconds) measured as function of request size 
	(number of points $N_p$), when the $N_p$ points are arranged  along a single line with 
	a separation between points of $2\pi/N_p$. Different symbols are the same as in Fig. 
	\ref{fig-timesa}.\label{fig-timesb}}
        \end{figure}

\begin{figure} [ht]
        \centering
        \includegraphics[width=0.5\linewidth]{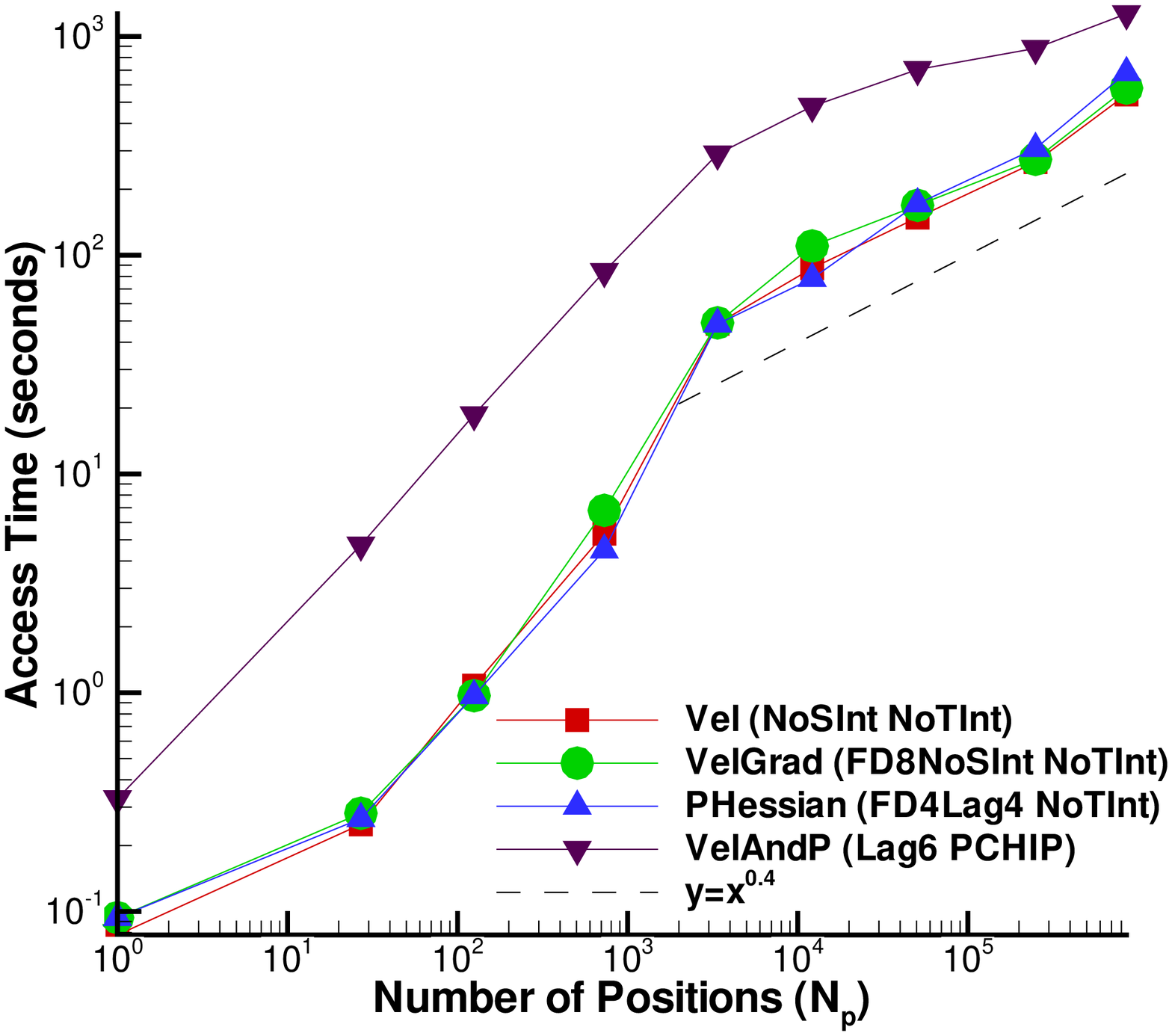}
        \caption{Access times (in seconds) measured as function of request size (number of points $N_p$), 
	when the $N_p$ points are distributed randomly in the entire dataset. 
	Different symbols are the same as in Fig. \ref{fig-timesa}. \label{fig-timesc}}
        \end{figure}

        To provide a more complete characterization of the presently achievable access speeds as
        function of the size of the request,
        additional experiments are carried out. Calls to the database are issued
        over the Internet  from a  FORTRAN program running on a desktop computer at JHU. Access times are       
        measured as  function of the number of points, $N_p$, at which data are requested.  Three spatial
        arrangements of the points are considered: 
         (a) $N_p$ points arranged on a cubic lattice, with $\sim N_p^{1/3}$ points on 
        each side; (b) all $N_p$ points along a single line, with points separated by $2\pi/N_p$; and 
	(c) $N_p$  positions  at random positions over the entire domain (uniform probability density 
	for each coordinate between $0$ and $2\pi$).  Figures \ref{fig-timesa}, 
	\ref{fig-timesb}  and \ref{fig-timesc}   show the resulting access times in seconds, 
	as function of $N_p$ for the three cases, respectively.  
	For each spatial arrangement of the points, four types of data are requested. 
	(1) three components of velocity (using \verb=GetVelocity=), without  spatial nor temporal interpolation; 
	(2) nine velocity gradient tensor elements (using \verb=GetVelocityGradient=),  
	evaluated using 8th-order centered finite differencing but without spatial nor temporal interpolation; 
	(3) pressure Hessian (using \verb=GetPressureHessian=), evaluated using 4th-order centered finite differencing 
	and 4th-order Lagrange Polynomial interpolation in space, by without temporal interpolation; 
	and finally (4) velocity vector and pressure (using \verb=GetVelocityAndPressure=), 
	with 6th-order Lagrange Polynomial interpolation in space and Piecewise Cubic Hermite 
	Interpolation Polynomial method for time. The corresponding access times as function of 
	$N_p$ are indicated with different symbols in Figs. \ref{fig-timesa}-\ref{fig-timesc}.  

In case (a), not surprisingly, the function evaluation using both spatial and 
temporal interpolation is the most time-consuming while the one without any interpolation is fastest. 
In this case, when the number of points requested increases, the access time increases as 
$\sim N_p^{0.9}$, i.e. almost linearly. In cases (b) and (c), there is not as 
much difference as in case (a) among the function evaluations without temporal interpolation, 
but the one with temporal interpolation still takes more access time. 
In the latter two cases, the access time increases as  $\sim N_p^{0.4}$ for $N_p$ above $\sim 3000$. 
Comparing the three cases, we can observe that when points are arranged as cubes, it takes the 
least access time for moderate size requests. The reason is that much of the data then 
typically falls within only a few of the stored atoms, decreasing the amount of data 
that needs to be read from disks.  As can be seen, access times depend upon 
the details of the data arrangement, interpolation type and the size of the request. 
Of course, 
it also depends on the client's network speed, etc. 

        \section{Analysis of the advected delta-vee system\label{sect:deltavee}}

        In this section, the database is used to analyze the advected delta-vee system
        derived in Ref. \cite{LiMeneveau05}. For the convenience of the readers, the derivation of
        the system is briefly reviewed first. We restrict ourselves to dynamics in three dimensional space.

        The advected delta-vee system decribes the Lagrangian evolution of the longitudinal
        and transverse velocity increments 
        (or more precisely, the longitudinal and transverse velocity gradient multiplied by a small 
        and fixed displacement).
        Specifically, for an incompressible velocity
        field $\ovl{u}_i({\bf x},t)$ filtered at scale $\Delta$ and
        velocity gradient $\overline{A}_{ij}\equiv
        \partial \overline{u}_i/\partial x_j$, consider a displacement
        vector ${\bf {r}}(t)$, and the unit vector in its direction
        $\hat{\bf r}={\bf r}/r$. The longitudinal and the magnitude of the
        transverse components of the `velocity increment' vector along this
        direction across a fixed distance $\ell<\Delta$ are defined as
        \be
        \delta u \equiv \ell~ \overline{A}_{rr} \equiv \ell~
        \overline{A}_{ik}~ \hat{r}_k~\hat{r}_i, \quad \delta v
        \equiv \ell \left|P_{ij}({\bf r})~\overline{A}_{jk}~ \hat{r}_k \right|,
        \label{eq:defduv}
        \ee
        where $\overline{A}_{rr}\equiv \overline{A}_{ik}
        \hat{r}_k\hat{r}_i$ is the velocity gradient along the direction
        $\hat{\bf r}$, and $P_{ij}({\bf r})\equiv
        \delta_{ij}-\hat{r}_i\hat{r}_j$ is the projection operator. For a detailed sketch
        illustrating these definitions, see Fig. 1 of  Ref.  \cite{LiMeneveau05}.

        Using the equations for the line element $r_i$ and the
        filtered velocity gradient $\ovl{A}_{ij}$, one can derive from the
        definitions of $\delta u$ and $\delta v$ the following equations
           (for details see \cite{LiMeneveau05, LiMeneveau06}):
        \bea
          \frac{d \delta u}{dt}&=&   -{\frac{1}{3} \delta u^2}~{\ell}^{-1} +{ \delta
        v^2}~{\ell}^{-1} + Q^- \ell  + Y,\label{eq:dudt0} \\
        \frac{d \delta v}{dt}& =&  - {2~\delta u~ \delta v}~{\ell}^{-1} +
        Z,\label{eq:dvdt0}
        \eea
        where $d/dt$ indicates material derivative. Also, $Y=\ell H_{ij}\hat{r}_i
        \hat{r}_j$ and $Z=\ell H_{ij} \hat{r}_j \hat{e}_i$, in which ${\bf\hat{e}}$
        is a unit vector in the direction of the transverse velocity-increment
        component (perpendicular to $\hat{\bf r}$) and $
     H_{ij}=-(\partial^2_{ij}\overline{p}-\frac{1}{3}\delta_{ij}\partial^2_{kk}\overline{p})
        -(\partial^2_{jk}\tau_{ik}-\frac{1}{3}\delta_{ij}\partial^2_{lk}\tau_{lk})
        +\nu\partial^2_{kk}\overline{A}_{ij} + (\ptl_j \ovl{f}_i - \frac{1}{3}\delta_{ij} \ptl_k \ovl{f}_k) $.
        $H_{ij}$ is the anisotropic part of the gradients of
        pressure, sub-grid scale, viscous, and external forces,
        with $\tau_{ij}\equiv\overline{u_i u}_j -\overline{u}_i\overline{u}_j$
        being the sub-grid scale (SGS) stress.  $Q^-$ is defined as  
		$Q^-= -\ovl{A}_{ij}\ovl{A}_{ji}/3 - 2 \delta u^2/3\ell^2$
        (see \cite{LiMeneveau06} for details).

        Eqs. \ref{eq:dudt0} and \ref{eq:dvdt0} are called 
		the ``advected delta-vee system''. In the system, the nonlinear
        self-interaction terms are closed, while $Q^-$, $Y$ and $Z$ are not. As 
        shown in \cite{LiMeneveau06}, the truncated system retaining only the nonlinear self-interaction
        terms is already able to predict a number of well-known, important intermittency trends. 
        Nevertheless, the unclosed terms have
        to be included through physically realistic models in order for the system to predict quantitative
        features of intermittency and to reach stationary statistical states. As the first step in constructing such models, 
        here we will now analyze the unclosed terms using the DNS data in the database.
        Since filtering operations are not yet supported by the database,
        the analysis is restricted to unfiltered DNS data and the results pertain to velocity increments across distances
        on the order of the Kolmogorov scale. Therefore the term corresponding
        to SGS force in $H_{ij}$ will be absent, and the filtered quantities will recover their
        unfiltered values. Hence we will drop in the following the overbar in the symbols. For example,
	$\overline{A}_{ij}$ will be replaced by $A_{ij}$ and so on. 

        The analysis is based on the evolution equation for the joint PDF of
        $\delta u $ and $\delta v$, which has been derived in appendix
        \ref{sect:jpdfeq}:
        \bea \label{eq:jpdf}
        & &\frac{\ptl P}{\ptl t}+\frac{\ptl}{\ptl \delta u}\left[(\delta
        v^2-\frac{1}{3}\delta u^2)P\right]+\frac{\ptl}{\ptl \delta v}
        \left[-2\delta u \delta v P\right]
        +\frac{\ptl }{\ptl \delta u}\left[\lla\left.Q^-\right|\delta u,\delta v\rra P\right]\nonumber\\
        &+&\frac{\ptl}{\ptl\delta u}\left[\lla\left.Y\right|\delta u,\delta
        v\rra P\right] +\frac{\ptl}{\ptl\delta
        v}\left[\lla\left.Z\right|\delta u,\delta v\rra P\right] + 3(\delta
        u-\lal\delta u\ral)P=0.
        \eea
        In Eq. \ref{eq:jpdf}, $P\equiv P(\delta u,\delta v;t)$ is the joint
        PDF of $\delta u$ and $\delta v$ at time $t$. The equation describes the balance produced by
        three different types of terms: the unsteady term
        $\ptl P/\ptl t$, the measure correction term ${\rm S}\equiv 3(\delta
        u-\lal\delta u\ral)P$, and the probability flux terms. For stationary turbulence,
        the unsteady term is zero. To simplify
        later analysis, we then write $\vct{\nabla}\cdot{\bf W} +S=0$, with ${\bf W}= {\bf W}_{\rm sc}+
       {\bf W}_{\rm Q}+ {\bf W}_{\rm p}+ {\bf W}_{\rm v}+{\bf W}_{\rm f}$. The following symbols  denote the
        probability fluxes caused by different physical mechanisms:
        \be \label{eq:w}
        \begin{array}{ll}
        {\bf W}_{\rm sc}=[\delta v^2-\delta u^2/3,-2\delta u \delta v]P, &
        {\bf W}_{\rm Q}=[\lla\left.Q^-\right|\delta u,\delta v\rra,0]P,\\
        {\bf W}_{\rm p}=[\lla\left. Y_p\right|\delta u,\delta v\rra,\lla
        \left. Z_p\right|\delta u,\delta v\rra]P,&
        {\bf W}_{\rm v}=[\lla\left. Y_v\right|\delta u,\delta v\rra,\lla \left. Z_v\right|\delta u,\delta v\rra]P,\\
        {\bf W}_{\rm f}=[\lla\left. Y_f\right|\delta u,\delta v\rra,\lla
        \left. Z_f\right|\delta u,\delta v\rra]P,&
        \\
        \end{array}
        \ee
        in which $Y$ and $Z$ have each been separated into three parts:
        \be \label{eq:yz}
        \begin{array}{ll}
        \displaystyle{Y_p=
        -\hat{r}_i\hat{r}_j\ptl^2_{ij}p+(1/3)\ptl^2_{kk}p},&
        Z_p= -\hat{e}_i\hat{r}_j\ptl^2_{ij}p,\\
        \displaystyle{Y_v= \nu \hat{r}_i \hat{r}_j \ptl^2_{kk}A_{ij}},&
        Z_v= \nu \hat{e}_i \hat{r}_j \ptl^2_{kk}A_{ij},\\
        \displaystyle{Y_f= \hat{r}_i\hat{r}_j\ptl_j f_i - (1/3)\ptl_k f_k},
        & Z_f = \hat{e}_i\hat{r}_j\ptl_j f_i.
        \end{array}
        \ee
        These terms represent, respectively, the effects of pressure, viscous diffusion, and external force.
        Note that in the $Z$'s the contributions from the isotropic parts of
        the tensors are zero because $\hat{\bf r}$ is orthogonal to
        $\hat{\bf e}$. The same is true in $Y_v$ due to incompressibility.

        \begin{figure}[ht]
        \center
        \includegraphics[width=0.6\linewidth]{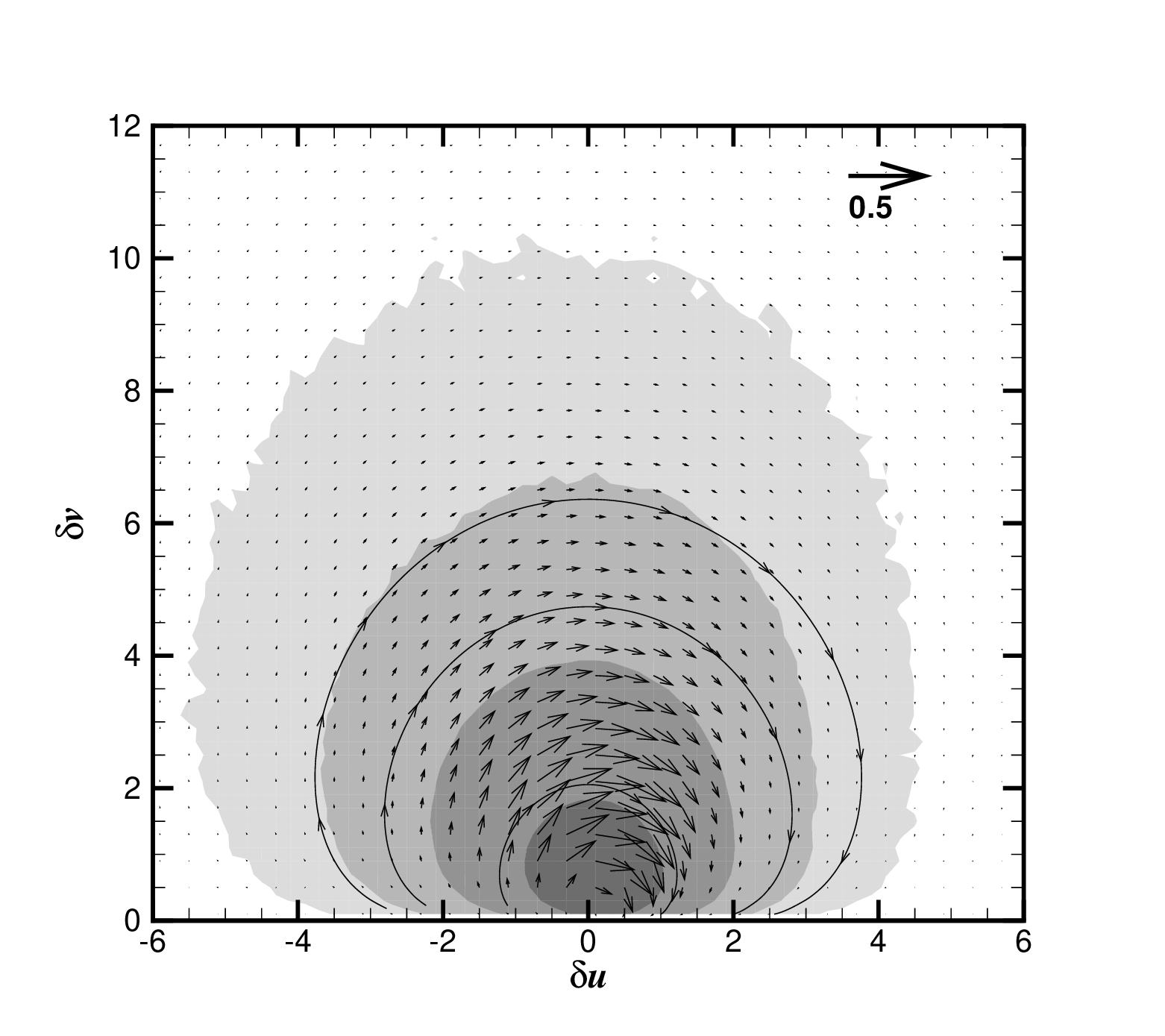}
        \caption{\label{fig:re_db} The vector plot of ${\bf W}_{\rm sc}$,
        the probability flux generated by the self-interaction terms.
        Several stream traces are plotted. The contour plot is $P(\delta u,
        \delta v)$, the joint PDF of $\delta u$ and $\delta v$.}
        \end{figure}
        \begin{figure}[ht]
        \center
        \includegraphics[width=0.6\linewidth]{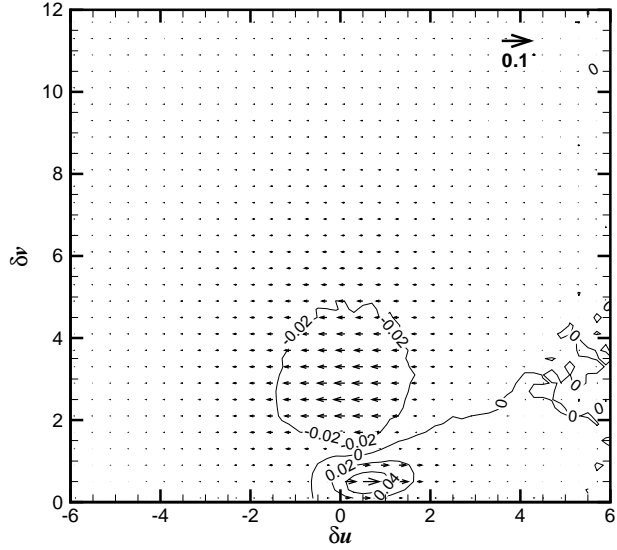}
        \caption{\label{fig:q_db} The vector plot of ${\bf W}_{\rm Q}$, the
        probability flux generated by the $Q^-$ term. By definition, all the
        vectors are parallel to $\delta u$ axis. Several contours of the
        magnitude of the vectors are plotted. }
        \end{figure}
        \begin{figure}[ht]
        \center
        \includegraphics[width=0.6\linewidth]{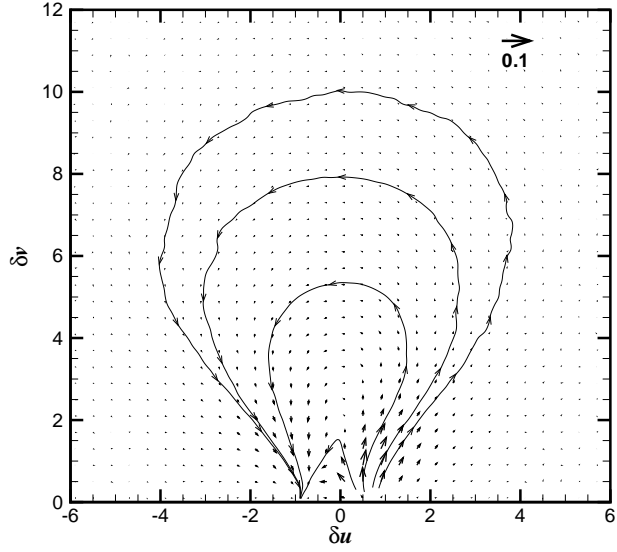}
        \caption{\label{fig:p_db} The vector plot of ${\bf W}_{\rm p}$, the
        probability flux generated by the anisotropic part of the pressure
        Hessian. }
        \end{figure}
        \begin{figure}[ht]
        \center
        \includegraphics[width=0.6\linewidth]{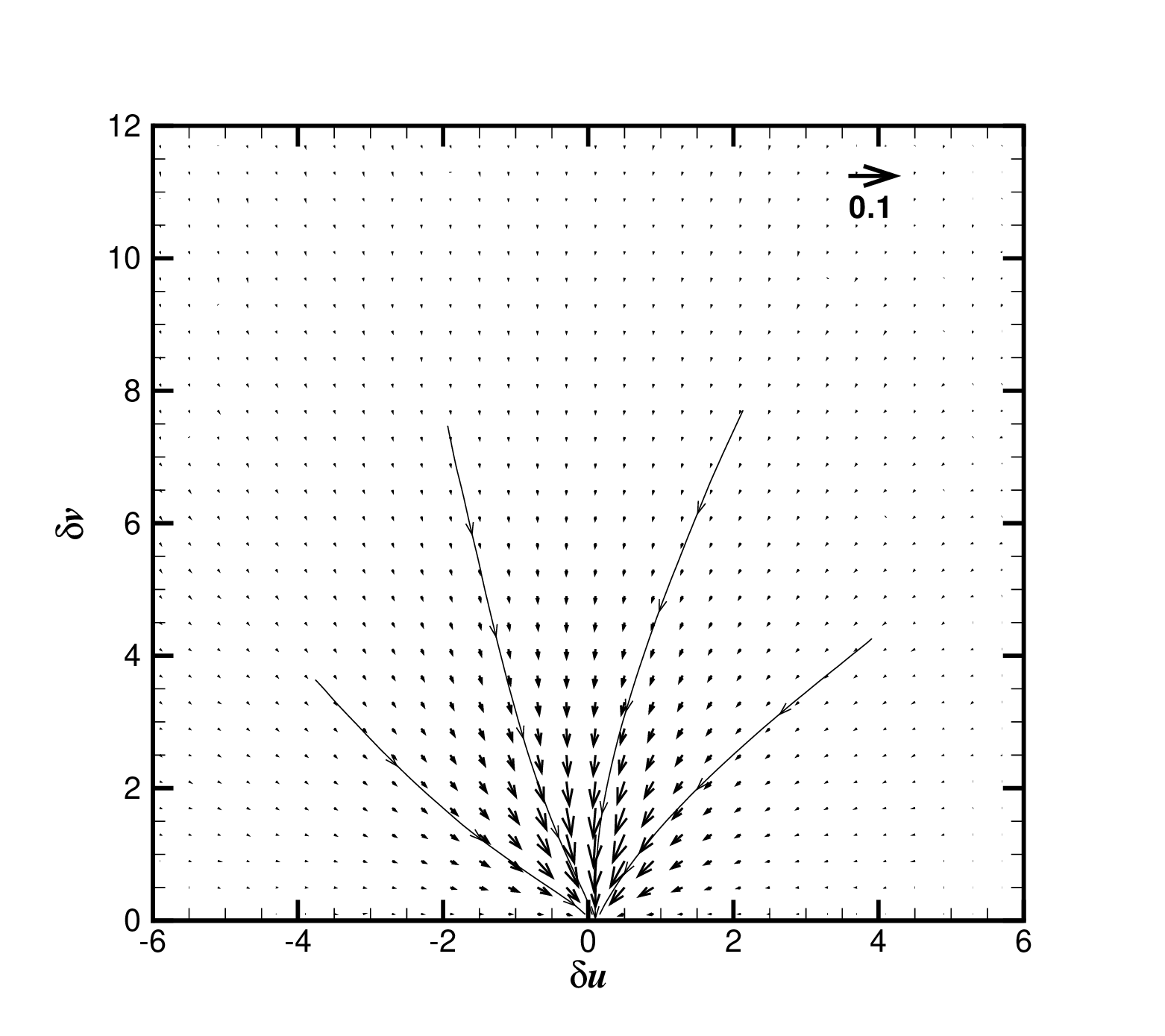}
        \caption{\label{fig:v_db} The vector plot of ${\bf W}_{\rm v}$, the
        probability flux generated by the viscous diffusion term.}
        \end{figure}

        The joint pdf $P(\delta u, \delta v)$ and the flux terms
        in the above joint PDF equation are calculated using the turbulence
        database described in previous sections, except for the flux
        produced by the forcing term, which is not included in the analysis.
        The forcing term can not, at this stage, be obtained directly from the database. But since the forcing only   
        occurs at large scales its direct effect on the terms related to the small-scale dynamics of
        velocity gradient are negligible, and its omission does
        not alter the conclusions to be drawn. Also, when
		calculating the viscous diffusion term ${\bf W}_{\rm v}$, $\nabla^2 A_{ij}$ is
		needed, which entails third
        order differentiation of the velocity field. This is a
        rather specific operation and, for now, was not considered generic enough to be
        included in the processing function set of the database.
		Therefore, in this analysis  $\nabla^2 A_{ij}$ is calculated on the local client-side computer 
		using the data of $A_{ij}$ obtained from database queries at various grid points, and 
		then evaluating the $4$th order
		central finite difference approximation to the Laplacian $\nabla^2$. 
          The $(\delta u,\delta v$) phase-space is binned into $80\times 80$ boxes
          and conditional averages are computed by averaging over about half of a million 
		  randomly chosen datapoints, and $50$ random directions of $\hat{\bf r}$ at $t=0$.

        The probability flux generated by the closed self-interaction terms
        is plotted in Fig. \ref{fig:re_db}. The contour plot layered under
        the vector plot is the joint PDF of $\delta u$ and $\delta v$. The
        joint PDF displays the correct skewness towards
        negative $\delta u$ direction. By construction, the stream traces in Fig.
        \ref{fig:re_db} have the same shapes as the phase orbits of the truncated
		system (with $Q^-$, $Y$, and $Z$ neglected), as plotted in
		Fig. 5(b) in \cite{LiMeneveau06}. As explained in \cite{LiMeneveau05, LiMeneveau06}, 
		due to the negative quadratic term $-\delta u^2 /3$
		in Eq. \ref{eq:dudt0}, negative values of $\delta u$ are continuously amplified, while
		positive fluctuations tend to be decreased. This mechanism,
		referred to as the self-amplification of the
		longitudinal velocity increments, is thus responsible for generating the negative 
		skewness in $\delta u$. On the other hand, 
		the right-hand side of Eq. \ref{eq:dvdt0} becomes positive
		when $\delta u$ is negative. Therefore, a large negative value of $\delta u$ will produce 
		exponential growth in $\delta v$, thus leading to strong fluctuation in $\delta v$. This 
		mechanism for the generation of intermittency in the transverse velocity increment is called
		the cross amplification mechanism. 
		The stream traces in Fig. \ref{fig:re_db} give an instructive 
		illustration of the coupling between the two
		mechanisms. 
		 This figure, however, also shows that the
        self-interaction terms alone can not maintain a stationary
        distribution in $P(\delta u, \delta v)$. We expect that the other
        terms will tend to oppose the excessive growth of the intermittency
        produced by these two terms and enable the system to reach a
        stationary distribution.

        Plotted in Fig. \ref{fig:q_db} through Fig. \ref{fig:v_db} are the
        probability fluxes generated by the $Q^-$ term, the anisotropic
        pressure Hessian term, and the viscous diffusion term, respectively.
        By definition, the $\delta v$ component of the flux ${\bf W}_{\rm
        Q}$ is zero, therefore all the vectors in Fig. \ref{fig:q_db} are
        parallel to the $\delta u$ axis. The lines in the figure are
        contours of the distribution of the non-zero (horizontal)  component of the
        probability flux vector. Separated by the zero contour, the vectors in the lower
        right corner point to positive $\delta u$, while those above point
        to the negative direction. This combination shows a general tendency
        for the $Q^-$ term to oppose the effect of the self-interactions
        terms. Note however, in the negative $\delta u$ half plane, the
        vectors pointing to the negative $\delta u$ direction would
        strengthen the negative amplification of $\delta u$, and thus increase
        the intermittency in $\delta v$ due to the cross-amplification
        mechanism explained before. On the other hand, the stream traces
        of the flux generated by the pressure Hessian, shown in Fig.
        \ref{fig:p_db}, have a similar shape as those in Fig.
        \ref{fig:re_db}, but with opposite direction. It indicates that the
        main effect of the pressure Hessian is to reduce the intermittency
        generated by the self-interaction terms. Besides, we note that the
        stream traces show an interesting sink-source structure. Fig.
        \ref{fig:v_db} shows that the contribution of viscous diffusion is
        mainly to introduce a nearly linear damping effect, i.e., all vectors
        pointing to the origin.

\section{Conclusions\label{sect:conclusions}}

In this paper, a new turbulence database system has been described. By
applying database technologies, the system enables remote access to a DNS 
data set with 27 Terabytes, encapsulating a complete
$1024^4$ space-time history of forced isotropic turbulence. The architecture of
the system, including the user interface, processing function set, data indexing and
partition, and work load scheduling, is explained. Test cases are
presented to demonstrate the usage of the system, and to check
against possible errors.  The database thus provides a new platform for
turbulence research, and a useful prototype for building large-scale
scientific databases.

As an example for the application of the database, the advected delta-vee system
introduced in \cite{LiMeneveau05, LiMeneveau06} is analyzed by accessing the
database from a remote desktop computer. The
analysis shows that the main effect of the unclosed terms in the
system is to oppose the excessive growth of intermittency in the
velocity increments produced by the nonlinear self-interaction
terms. The various terms (isotropic pressure Hessian effect,
deviatoric pressure Hessian term, and viscous term) show markedly different behaviors
among themselves. While the detailed physics underlying these observations are not yet clear,
the results imply that when closure models are constructed for
these terms, they should be modeled separately. Specifically, the viscous term may be modeled
using a simple relaxation towards the origin, but the pressure terms will require more elaborate closures.

\section*{Acknowledgments}
The authors thank Tamas Budavari, Ani Thakar, Jan Vandenberg, and Alainna Wonders 
for their valuable help at various stages of development and maintenance of the database cluster. 
They also wish to acknowledge the National Science Foundation for funding through the ITR
program, grant AST-0428325. The DNS was performed on a cluster supported by NSF through
MRI grant CTS-0320907. Additional hardware support was provided by the Moore Foundation. Discussions
with Prof. Ethan Vishniac are also gratefully acknowledged.

\section*{Appendices}
\appendix
\section{Equation for the joint PDF of the
longitudinal and transverse velocity increments\label{sect:jpdfeq}}

In this appendix, we will derive the equation for the joint PDF of
$\delta u$ and $\delta v$, from the dynamic equations for $\delta u
$ and $\delta v$ (Eqs. \ref{eq:dudt0} and \ref{eq:dvdt0}), and the equation for
the length of the line element $r(t)$:
\be \label{eq:drdt}
 \frac{dr}{dt}=\delta u ~ r ~ \ell^{-1}.
\ee
Recall that the velocity increments are defined over a distance $\ell$ along
the directions of evolving line elements $\bf r$. As noted in \cite{LiMeneveau05, LiMeneveau06},
during their evolution the line elements tend to concentrate along
the stretching directions of the flow field. Therefore, the joint PDF defined with
equal weight on each {\it trajectory} in the $(\delta u, \delta v)$
phase space will put more weight on the velocity increments along
the stretching directions. In order to obtain unbiased statistics, a
measure correcting procedure has been proposed in \cite{LiMeneveau05}.
Based on mass conservation, it was shown that when accumulating the statistics
of $(\delta u, \delta v)$ at time $t$, each trajectory has to be
weighed with a factor $[r_0/r(t)]^3$, where $r_0$ is the initial
length of the line element. In
terms of the joint PDF of $\delta u$ and $\delta v$, it implies
that, assuming all the line elements initially having the same
length $\ell$, the unbiased joint PDF of $\delta u$ and $\delta v$
can be formally written as
\be
P(\delta u, \delta v; t)=\frac{1}{I(t)}\int_0^{+\infty} P_3(\delta
u,\delta v, r; t) r^{-3} dr,
\ee
in which $P_3(\delta u, \delta v, r)$ is the traditional joint PDF
of the three independent variables. $I(t)$ is a normalization
factor, defined as:
\be
I(t)\equiv\int_{-\infty}^{+\infty}\int_0^{+\infty}\int_0^{+\infty}P_3(\delta
u, \delta v, r; t) r^{-3} d\delta u\, d\delta v\, dr.
\ee

From these definitions, the equation for $P$ can be derived from
the Liouville equation for $P_3$
\be \label{eq:p3}
\frac{\ptl P_3}{\ptl t} +\frac{\ptl}{\ptl \delta
u}(\lla\left.\delta\dot{u}\right|\delta u,\delta v,r\rra P_3)+
\frac{\ptl}{\ptl\delta v}(\lla\left.\delta\dot{v}\right|\delta
u,\delta v,r\rra P_3) +\frac{\ptl}{\ptl r}(\lla\left.
\dot{r}\right|\delta u,\delta v,r\rra P_3)=0,
\ee
where $\dot{a}\equiv da/dt$ means material derivative.
Multiplying Eq. \ref{eq:p3} by $r^{-3}$ and integrating over $r$,
one obtains
\bea
\frac{\ptl I(t)P}{\ptl t} \!\!&+& \!\! \frac{\ptl}{\ptl\delta
u}\left[\int_0^{+\infty}\lla\left. \delta\dot{u}\right|\delta
u,\delta v,r\rra P_3r^{-3}dr\right]+ \frac{\ptl}{\ptl \delta
v}\left[\int_0^{+\infty}\lla\left.
\delta \dot{v}\right|\delta u,\delta v,r\rra P_3r^{-3}dr\right]\nonumber\\
\!\!&+&\!\! \int_0^{+\infty}r^{-3}\frac{\ptl}{\ptl r}(\lla\left.
\dot{r}\right|\delta u, \delta v,r\rra P_3 ) dr = 0.
\eea
By straightforward calculation, the first two integrals can be reduced to:
\be
I(t)\lla \left.\delta \dot{u}\right|\delta u, \delta v\rra P(\delta
u,\delta v)
\ee
and
\be
I(t)\lla\left.\delta\dot{v}\right| \delta u,\delta v\rra P(\delta
u,\delta v)
\ee
respectively. Using Eq. \ref{eq:drdt}, the third
one becomes
\be
\int_0^{+\infty}r^{-3}\frac{\ptl}{\ptl r} (\delta u r P_3) dr =
r^{-2}\delta uP_3|_0^{+\infty}-\int_0^{+\infty}\delta
urP_3(-3)r^{-4}dr=3\delta uI(t)P,
\ee
assuming $\delta uP_3(\delta u,\delta v,r)$ approaches zero faster
than $r^{2}$ for any $\delta u$, $\delta v$ when $r\to 0$. Thus,
one obtains
\be
\frac{\ptl I(t)P}{\ptl t}+ I(t)\left[\frac{\ptl}{\ptl \delta u}(\lla
\left.\delta \dot{u}\right|\delta u,\delta v\rra P)
+\frac{\ptl}{\ptl \delta v}(\lla \left. \delta \dot{v}\right|\delta
u,\delta v\rra P)+3\delta uP\right]=0.
\ee
Integrating the above equation over $\delta u$ and $\delta v$, the
normalization condition for $P$ yields
\be
\frac{d
I(t)}{dt}+3I(t)\int_{-\infty}^{+\infty}\int_0^{+\infty}\delta uP
d\delta ud\delta v=\frac{dI(t)}{dt} +3I(t)\lal \delta u\ral=0,
\ee
if $\lla\left.\delta \dot{u}\right|\delta u,\delta v\rra P$ and
$\lla\left.\delta \dot{v}\right|\delta u,\delta v\rra P$ tend to
zero at the boundaries. Thus
\bea
\frac{\ptl P}{\ptl t}&=&\frac{1}{I(t)}\frac{\ptl I(t)P}{\ptl t}-\frac{P}{I(t)}\frac{d I(t)}{dt}\nonumber\\
&=&-\left[\frac{\ptl}{\ptl \delta u}(\lla \left.\delta
\dot{u}\right|\delta u,\delta v\rra P) +\frac{\ptl}{\ptl \delta
v}(\lla \left. \delta \dot{v}\right|\delta u,\delta v\rra P)+3\delta
uP\right]\nonumber\\
& &+3\lal \delta u\ral P.
\eea
Replacing $\delta \dot{u}$ and $\delta \dot{v}$ with Eq.
\ref{eq:dudt0} and \ref{eq:dvdt0} and rearranging the terms finally
yields
\bea \label{eq:jpdfa}
& &\frac{\ptl P}{\ptl t}+\frac{\ptl}{\ptl \delta u}\left[\left(\delta
v^2-\frac{1}{3}\delta u^2\right)P\right]+\frac{\ptl}{\ptl \delta v}
\left[-2\delta u \delta v P\right]
+\frac{\ptl }{\ptl \delta u}\left[\lla\left. Q^-\right|\delta u,\delta v\rra P\right]\nonumber\\
&+&\frac{\ptl}{\ptl\delta u}\left[\lla\left.Y\right|\delta u,\delta
v\rra P\right] +\frac{\ptl}{\ptl\delta
v}\left[\lla\left.Z\right|\delta u,\delta v\rra P\right] + 3(\delta
u-\lal\delta u\ral)P=0.
\eea
This is the equation for the joint PDF of $\delta u$ and $\delta v$.
The last term on the LHS of the equation comes from the measure
correction procedure. Note that we have assumed $\ell = 1$.

\section{Documentation: database spatial differentiation}

In this appendix, details are provided about the various options for spatial differentiations 
implemented in the database cluster.  Below, $f$ denotes any one of the three components of
velocity, $u$, $v$ or $w$, or pressure, $p$, depending on which
function is called. $\Delta x$ and $\Delta y$ are the width of
grid in $x$ and $y$ direction.

\subsection{Options for GetVelocityGradient and GetPressureGradient}

\begin{figure}[h]
\centering\includegraphics[width=1.0\linewidth]{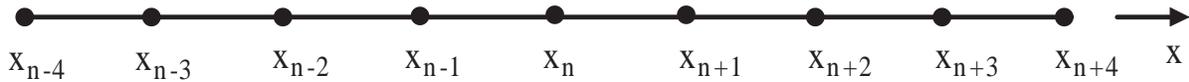}
\caption{Illustration of data points along $x$ direction. 
The same approach is used in the $y$ and $z$ directions.}\label{1DGrid}
\end{figure}

\subsubsection*{FD4: 4th-order centered finite differencing}
\label{sec-FD4}

With the edge replication of 4 data-points on each side, 
this option can be spatially interpolated using 4th-order Lagrange Polynomial interpolation.

\bea \left.\frac{df}{dx}\right|_{x_n}
&=& \frac{2}{3\Delta x}[f(x_{n+1})-f(x_{n-1})]-\frac{1}{12\Delta
x}[f(x_{n+2})-f(x_{n-2})]\nonumber\\
&&+O(\Delta x^4) \eea

\subsubsection*{FD6: 6th-order centered finite differencing}

\bea \left.\frac{df}{dx}\right|_{x_n} &=&
\frac{3}{4\Delta
x}[f(x_{n+1})-f(x_{n-1})]-\frac{3}{20\Delta x}[f(x_{n+2})-f(x_{n-2})]\nonumber\\
&&+\frac{1}{60\Delta x}[f(x_{n+3})-f(x_{n-3})]+O(\Delta x^6) \eea

\subsubsection*{FD8: 8th-order centered finite differencing}
With the edge replication of 4 data-points on each side, this is the highest-order finite difference option available.
\bea \left.\frac{df}{dx}\right|_{x_n} &=&
\frac{4}{5\Delta
x}[f(x_{n+1})-f(x_{n-1})]-\frac{1}{5\Delta x}[f(x_{n+2})-f(x_{n-2})]\nonumber\\
&&+\frac{4}{105\Delta x}[f(x_{n+3})-f(x_{n-3})]-\frac{1}{280\Delta
x}[f(x_{n+4})-f(x_{n-4})]\nonumber\\
&& +O(\Delta x^8) \eea

\subsection{Options for GetVelocityLaplacian, GetVelocityHessian and
GetPressureHessian}
In this section, second derivatives finite difference evaluations are shown. The expressions are
given for derivatives along single directions in terms of the $x$-direction, and mixed derivatives are
illustrated on the $x$-$y$ plane. The same approach is used in the $y$ and $z$ 
directions, as well as in the $x$-$z$ and $y$-$z$ planes
for the other mixed derivatives.

\subsubsection*{FD4: 4th-order centered finite differencing (can be spatially
interpolated using 4th-order Lagrange Polynomial interpolation)}
\bea \left.\frac{d^2f}{dx^2}\right|_{(x_m,y_n)} &=&
\frac{4}{3\Delta
x^2}[f(x_{m+1},y_n)+f(x_{m-1},y_n)-2f(x_m,y_n)]\nonumber\\
&&-\frac{1}{12\Delta
x^2}[f(x_{m+2},y_n)+f(x_{m-2},y_n)-2f(x_m,y_n)]\nonumber\\
&&+O(\Delta x^4) \eea

\bea \left.\frac{d^2f}{dxdy}\right|_{(x_m,y_n)} &=&
\frac{1}{3\Delta
x\Delta y}[f(x_{m+1},y_{n+1})+f(x_{m-1},y_{n-1})\nonumber\\
&&-f(x_{m+1},y_{n-1})-f(x_{m-1},y_{n+1})]\nonumber\\
&&-\frac{1}{48\Delta
x\Delta y}[f(x_{m+2},y_{n+2})+f(x_{m-2},y_{n-2})\nonumber\\
&&-f(x_{m+2},y_{n-2})-f(x_{m-2},y_{n+2})]\nonumber\\
&&+O(\Delta x^4,\Delta y^4) 
\eea

\subsubsection*{FD6: 6th-order centered finite differencing}
\bea \left.\frac{d^2f}{dx^2}\right|_{(x_m,y_n)}
&=& \frac{3}{2\Delta
x^2}[f(x_{m+1},y_n)+f(x_{m-1},y_n)-2f(x_m,y_n)]\nonumber\\
&&-\frac{3}{20\Delta
x^2}[f(x_{m+2},y_n)+f(x_{m-2},y_n)-2f(x_m,y_n)]\nonumber\\
&&+\frac{1}{90\Delta
x^2}[f(x_{m+3},y_n)+f(x_{m-3},y_n)-2f(x_m,y_n)]\nonumber\\
&&+O(\Delta x^6) \eea

\bea \left.\frac{d^2f}{dxdy}\right|_{(x_m,y_n)} &=&
\frac{3}{8\Delta
x\Delta y}[f(x_{m+1},y_{n+1})+f(x_{m-1},y_{n-1})\nonumber\\
&&-f(x_{m+1},y_{n-1})-f(x_{m-1},y_{n+1})]\nonumber\\
&&-\frac{3}{80\Delta
x\Delta y}[f(x_{m+2},y_{n+2})+f(x_{m-2},y_{n-2})\nonumber\\
&&-f(x_{m+2},y_{n-2})-f(x_{m-2},y_{n+2})]\nonumber\\
&&+\frac{1}{360\Delta
x\Delta y}[f(x_{m+3},y_{n+3})+f(x_{m-3},y_{n-3})\nonumber\\
&&-f(x_{m+3},y_{n-3})-f(x_{m-3},y_{n+3})]\nonumber\\
&&+O(\Delta x^6, \Delta y^6) \eea

\subsubsection*{FD8: 8th-order centered finite differencing}
\bea \left.\frac{d^2f}{dx^2}\right|_{(x_m,y_n)}
&=& \frac{792}{591\Delta
x^2}[f(x_{m+1},y_n)+f(x_{m-1},y_n)-2f(x_m,y_n)]\nonumber\\
&&-\frac{207}{2955\Delta
x^2}[f(x_{m+2},y_n)+f(x_{m-2},y_n)-2f(x_m,y_n)]\nonumber\\
&&-\frac{104}{8865\Delta
x^2}[f(x_{m+3},y_n)+f(x_{m-3},y_n)-2f(x_m,y_n)]\nonumber\\
&&+\frac{9}{3152\Delta
x^2}[f(x_{m+4},y_n)+f(x_{m-4},y_n)-2f(x_m,y_n)]\nonumber\\
&&+O(\Delta x^8) \eea

\bea \left.\frac{d^2f}{dxdy}\right|_{(x_m,y_n)} &=&
\frac{14}{35\Delta
x\Delta y}[f(x_{m+1},y_{n+1})+f(x_{m-1},y_{n-1})\nonumber\\
&&-f(x_{m+1},y_{n-1})-f(x_{m-1},y_{n+1})]\nonumber\\
&&-\frac{1}{20\Delta
x\Delta y}[f(x_{m+2},y_{n+2})+f(x_{m-2},y_{n-2})\nonumber\\
&&-f(x_{m+2},y_{n-2})-f(x_{m-2},y_{n+2})]\nonumber\\
&&+\frac{2}{315\Delta
x\Delta y}[f(x_{m+3},y_{n+3})+f(x_{m-3},y_{n-3})\nonumber\\
&&-f(x_{m+3},y_{n-3})-f(x_{m-3},y_{n+3})]\nonumber\\
&&-\frac{1}{2240\Delta
x\Delta y}[f(x_{m+4},y_{n+4})+f(x_{m-4},y_{n-4})\nonumber\\
&&-f(x_{m+4},y_{n-4})-f(x_{m-4},y_{n+4})]\nonumber\\
&&+O(\Delta x^8,\Delta y^8) \eea

\section{Documentation: Database Spatial Interpolation Options}
In this appendix, details are provided about the various options for spatial interpolation 
implemented in the database cluster.  
\begin{figure}[h]
\centering\includegraphics[width=1.0\linewidth]{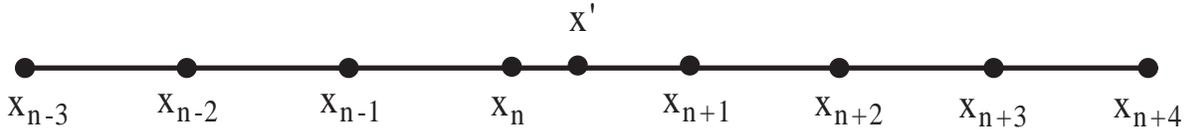}
\caption{Illustration of Lagrangian interpolation. }\label{Lagrangian}
\end{figure}
\subsection{Interpolation Options for GetVelocity and GetVelocityAndPressure}
In this section, $f$ denotes any one of the three components of
velocity, $u$, $v$ or $w$, or pressure, $p$, depending on which
function is called. $\Delta x$, $\Delta y$ and $\Delta z$ are the
width of grid in $x$, $y$ and $z$ direction. ${\bf
x'}=(x',y',z')$.
\subsubsection*{NoSInt: No spatial  interpolation}
\label{sec-noint}
In this case, the value at the datapoint closest to each
coordinate value is returned, rounding up or down in each
direction. \bea f({\bf x'}) &=& f(x_n,y_p,z_q)\eea where
$n=int(\frac{x'}{\Delta x}+\frac{1}{2})$, $p=int(\frac{y'}{\Delta
y}+\frac{1}{2})$, $q=int(\frac{z'}{\Delta z}+\frac{1}{2})$.

\subsubsection*{Lag4: 4th-order Lagrange Polynomial interpolation}

In this case, 4th-order Lagrange Polynomial interpolation is done along each spatial direction.
\bea f({\bf x'}) &=&
\sum_{i=1}^4\sum_{j=1}^4\sum_{k=1}^4f(x_{n-2+i},y_{p-2+j},z_{q-2+k})\nonumber\\
&&\cdot l_x^{n-2+i}(x')\cdot l_y^{p-2+j}(y')\cdot
l_z^{q-2+k}(z')\eea \bea l_\theta^i(\theta') &=&
\frac{\prod\limits_{j=n-1,j\neq
i}^{n+2}(\theta'-\theta_j)}{\prod\limits_{j=n-1,j\neq
i}^{n+2}(\theta_i-\theta_j)}\eea where $\theta$ can be $x$, $y$, or $z$.

\subsubsection*{Lag6: 6th-order Lagrange Polynomial interpolation}

In this case, 6th-order Lagrange Polynomial interpolation is done along each spatial direction.
\bea f({\bf x'}) &=&
\sum_{i=1}^6\sum_{j=1}^6\sum_{k=1}^6f(x_{n-3+i},y_{p-3+j},z_{q-3+k})\nonumber\\
&&\cdot l_x^{n-3+i}(x')\cdot l_y^{p-3+j}(y')\cdot
l_z^{q-3+k}(z')\eea \bea l_\theta^i(\theta') &=&
\frac{\prod\limits_{j=n-2,j\neq
i}^{n+3}(\theta'-\theta_j)}{\prod\limits_{j=n-2,j\neq
i}^{n+3}(\theta_i-\theta_j)}\eea where $\theta$ can be $x$, $y$, or $z$.

\subsubsection*{Lag8: 8th-order Lagrange Polynomial interpolation}

In this case, 8th-order Lagrange Polynomial interpolation is done along each spatial direction.
\bea f({\bf x'}) &=&
\sum_{i=1}^8\sum_{j=1}^8\sum_{k=1}^8f(x_{n-4+i},y_{p-4+j},z_{q-4+k})\nonumber\\
&&\cdot l_x^{n-4+i}(x')\cdot l_y^{p-4+j}(y')\cdot
l_z^{q-4+k}(z')\eea \bea l_\theta^i(\theta') &=&
\frac{\prod\limits_{j=n-3,j\neq
i}^{n+4}(\theta'-\theta_j)}{\prod\limits_{j=n-3,j\neq
i}^{n+4}(\theta_i-\theta_j)}\eea where $\theta$ can be $x$, $y$, or $z$.

\subsection{Interpolation Options for GetVelocityGradient, GetPressureGradient, GetVelocityLaplacian, GetVelocityHessian and GetPressureHessian}

In this section, $f$ denotes velocity or pressure gradient,
velocity or pressure Hessian, or velocity Laplacian, depending on
which function is called.
\subsubsection*{FD4NoInt, FD6NoInt, FD8NoInt: No spatial  interpolation}

In this case, the value of the 4th, 6th or 8th order finite-difference
evaluation of the derivative at the datapoint closest to each coordinate value is
returned, rounding up or down in each direction.
\bea f({\bf x'}) &=& f(x_n,y_p,z_q)\eea where
$n=int(\frac{x'}{\Delta x}+\frac{1}{2})$, $p=int(\frac{y'}{\Delta
y}+\frac{1}{2})$, $q=int(\frac{z'}{\Delta z}+\frac{1}{2})$.

\subsubsection*{FD4Lag4: 4th-order Lagrange Polynomial interpolation of the 4th-order finite differences}

In this case, the values of the 4th order finite-difference
evaluation of the derivative at the data points are interpolated using 4th-order Lagrange Polynomials.
\bea f({\bf x'}) &=&
\sum_{i=1}^4\sum_{j=1}^4\sum_{k=1}^4f(x_{n-2+i},y_{p-2+j},z_{q-2+k})\nonumber\\
&&\cdot l_x^{n-2+i}(x')\cdot l_y^{p-2+j}(y')\cdot
l_z^{q-2+k}(z')\eea \bea l_\theta^i(\theta') &=&
\frac{\prod\limits_{j=n-1,j\neq
i}^{n+2}(\theta'-\theta_j)}{\prod\limits_{j=n-1,j\neq
i}^{n+2}(\theta_i-\theta_j)}\eea where $\theta$ can be $x$, $y$, or $z$.

\section{Documentation: Database Temporal Interpolation Options}

\begin{figure}[h]
\begin{minipage}{\linewidth}
\centering\includegraphics[width=0.5\linewidth]{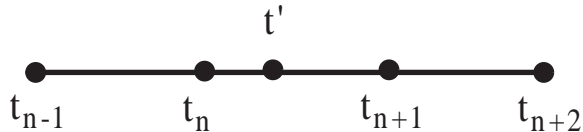}
\end{minipage} \caption{Illustration of data points for time.}\label{PCHIP}
\end{figure}

In this section, $f$ denotes velocity, pressure, velocity or
pressure gradient, velocity or pressure Hessian, or velocity
Laplacian, depending on which function is called. $\Delta t$ is
the time increment between consecutive times stored in the database.

\subsection*{NoTInt: No temporal interpolation}

In this case, the value at the datapoint closest to the time value
is returned, rounding up or down. \bea f(t') &=& f(t_n)\eea where
$n=int(\frac{t'}{\Delta t}+\frac{1}{2})$.

\subsection*{PCHIP: Cubic Hermite Interpolation in time}

The value from the two nearest time points is interpolated at time $t'$
using Cubic Hermite Interpolation Polynomial, with centered finite
difference evaluation of the end-point time derivatives - i.e. a
total of four temporal points are used. \bea
f(t')=a+b(t'-t_n)+c(t'-t_n)^2+d(t'-t_n)^2(t'-t_{n+1}) \eea where
\bea a &=& f(t_n)\nonumber\eea \bea b &=&
\frac{f(t_{n+1})-f(t_{n-1})}{2\Delta t}\nonumber\eea \bea c &=&
\frac{f(t_{n+1})-2f(t_n)+f(t_{n-1})}{2\Delta t^2}\nonumber\eea
\bea d &=&
\frac{-f(t_{n-1})+3f(t_n)-3f(t_{n+1})+f(t_{n+2})}{2\Delta t^3}
\nonumber\eea


\end{document}